\documentclass[a4paper,11pt]{article}
\pdfoutput=1 

\usepackage{amsmath}
\usepackage{amssymb}
\usepackage{amsthm}
\usepackage{graphicx}
\usepackage{slashed}
\usepackage{setspace}
\usepackage{floatflt}
\usepackage{multirow}

\newcommand{\bZ}{\mathbb{Z}}

\newcommand{\bR}{\mathbb{R}}

\newcommand{\cN}{\mathcal{N}}

\newcommand{\cO}{\mathcal{O}}

\newcommand{\cL}{\mathcal{L}}

\newcommand{\ov}{\overline}

\newcommand{\drawsquare}[2]{\hbox{%
\rule{#2pt}{#1pt}\hskip-#2pt
\rule{#1pt}{#2pt}\hskip-#1pt
\rule[#1pt]{#1pt}{#2pt}}\rule[#1pt]{#2pt}{#2pt}\hskip-#2pt
\rule{#2pt}{#1pt}}
\newcommand{\fund}{\raisebox{-.5pt}{\drawsquare{6.5}{0.4}}}
\newcommand{\Ysymm}{\raisebox{-.5pt}{\drawsquare{6.5}{0.4}}\hskip-0.4pt%
\raisebox{-.5pt}{\drawsquare{6.5}{0.4}}}
\newcommand{\Yasymm}{\raisebox{-3.5pt}{\drawsquare{6.5}{0.4}}\hskip-6.9pt%
\raisebox{3pt}{\drawsquare{6.5}{0.4}}}
\newcommand{\antifund}{\overline{\fund}}

\usepackage{color}

\usepackage{jheppub} 

\usepackage[T1]{fontenc} 

\title{String Consistency, Heavy Exotics, and the Diphoton Excess}

\title{String Consistency, Heavy Exotics, \\ and the $750$ GeV Diphoton Excess at the LHC}

\author[1,2]{Mirjam Cveti{\v c},}
\author[3]{James Halverson,}
\author[4,5]{and Paul Langacker}

\affiliation[1]{Department of Physics and Astronomy,  University of Pennsylvania, \\ Philadelphia, PA 19104-6396, USA}
\affiliation[2]{Center for Applied Mathematics and Theoretical Physics,  University of Maribor, \\ Maribor, Slovenia}
\affiliation[3]{Department of Physics, Northeastern University, \\ Boston, MA 02115-5000, USA}
\affiliation[4]{School of Natural Science, Institute for Advanced Study, \\ Einstein Drive, Princeton, NJ 08540, USA}
\affiliation[5]{Department of Physics, Princeton University\\ Princeton, NJ 08544, USA}

\emailAdd{cvetic@hep.upenn.edu}
\emailAdd{j.halverson@neu.edu}
\emailAdd{pgl@ias.edu}

\abstract{String consistency conditions are stronger than anomaly
  cancellation and can require the addition of exotics in the visible
  sector. We study such exotics and demonstrate that they may
  account for the modest excess at $750$ GeV in recent diphoton
  resonance searches performed by the ATLAS and CMS collaborations. In
  a previous analysis of type II MSSM D-brane quivers we
  systematically added up to five exotics for the sake of satisfying
  string consistency conditions. Using this dataset, we demonstrate
  that 89780 of the 89964 quivers have exotics, 78155 of which include
  singlets that may couple to MSSM or exotic multiplets with coupling
  structures governed by $U(1)$ symmetries that are often
  anomalous. We demonstrate that certain sets of exotics are far
  preferred over others and study the structure of singlet couplings
  to heavy exotics carrying standard model charges. Typical
  possibilities include singlets that may decay to vector-like quarks
  and / or vector-like leptons and subsequently to two photons. 
  We show that a narrow width diphoton
  excess can be accounted for while evading existing bounds if
  multiple exotics are added, with vector-like leptons of mass
  $M_L\lesssim 375$ GeV and vector-like quarks with masses up to $\simeq
  3$ TeV. However, a large width $(\Gamma/M \sim 0.06)$, as suggested
  by the ATLAS data, cannot be easily accommodated in this
  framework. Renormalization group equations with GUT-scale boundary
  conditions show that these supersymmetric models are perturbative
  and stable. Type IIA compactifications on toroidal orbifolds allow
  for $O(10)$ Yukawa couplings in the ultraviolet. We also
  discuss the possibility of accounting for the diphoton excess in a
  low string scale scenario via the decay of string axions. 
  
}

\begin{document}

\begin{flushright}
\parbox[t]{1.73in}{\flushright 
UPR-1276-T}
\end{flushright}
\maketitle
\flushbottom

\section{Introduction}
The ATLAS and CMS experiments have both seen \cite{Seminar,CMS,ATLAS} an excess in diphoton
final states at $750$ GeV using $13$ TeV data collected in
2015. 

Though the global significance of this excess is $< 3\sigma$, it
is interesting to ask whether there are simple models that could
explain the excess. Many models have been proposed \cite{Harigaya:2015ezk,Mambrini:2015wyu,Backovic:2015fnp,Pilaftsis:2015ycr,
Franceschini:2015kwy,Nakai:2015ptz,Buttazzo:2015txu,DiChiara:2015vdm,Higaki:2015jag,Knapen:2015dap,McDermott:2015sck,Ellis:2015oso,
Low:2015qep,Bellazzini:2015nxw,Gupta:2015zzs,
Petersson:2015mkr,Molinaro:2015cwg,Dutta:2015wqh,
Cao:2015pto,Matsuzaki:2015che,Kobakhidze:2015ldh,
Martinez:2015kmn,Cox:2015ckc,No:2015bsn,Demidov:2015zqn,Chao:2015ttq,
Fichet:2015vvy,Curtin:2015jcv,Bian:2015kjt,Chakrabortty:2015hff,Agrawal:2015dbf,Csaki:2015vek,
Ahmed:2015uqt,Falkowski:2015swt,Bai:2015nbs,Aloni:2015mxa,Gabrielli:2015dhk,Benbrik:2015fyz,Kim:2015ron,
Alves:2015jgx,Megias:2015ory,Carpenter:2015ucu,
Bernon:2015abk,Chao:2015nsm,
Arun:2015ubr,Han:2015cty,Chang:2015bzc,Chakraborty:2015jvs,Ding:2015rxx,Han:2015dlp,
Han:2015qqj,
Chang:2015sdy,Bardhan:2015hcr,Antipin:2015kgh,
Wang:2015kuj,
Cao:2015twy,Huang:2015evq,Heckman:2015kqk,Dhuria:2015ufo,Bi:2015uqd,Kim:2015ksf,
Berthier:2015vbb,Cho:2015nxy,Cline:2015msi,Bauer:2015boy,Chala:2015cev,
Barducci:2015gtd,Boucenna:2015pav,Murphy:2015kag,
Feng:2015wil,Hernandez:2015ywg,Pelaggi:2015knk,
Dey:2015bur,deBlas:2015hlv,Belyaev:2015hgo,Dev:2015isx} that can
account for the excess. (For an earlier study, see \cite{Jaeckel:2012yz}). These include models where the $750$ GeV particle
is a boson that decays to two photons via heavy charged fermions
running in a loop or directly via an axionic coupling to the
hypercharge field strength. 

It is also interesting to investigate whether
some of the models may be natural remnants of an ultraviolet
completion, such as string theory. (For an interpretation within the F-theory context see \cite{Heckman:2015kqk}.)
In this paper we will primarily focus on models in which visible sector
exotics must be added for the consistency of string compactifications;
for concreteness we will work in the context of type II orientifold
compactifications, though similar ideas also apply in other areas of
the landscape.  More specifically, chiral matter spectra in type II
compactifications are subject to string consistency conditions that go
beyond typical anomaly cancellation conditions, and nearly all
bottom-up realizations of the minimal supersymmetric standard model
(MSSM) in this context do not satisfy the constraints, even though the
MSSM is anomaly free. Such theories can be consistently embedded
in string theory only if exotic matter is added. These conditions not only
motivate the existence of exotics, but also dictate their
couplings.

In this sense, string consistency conditions provide a well-motivated
method for
expanding around the MSSM\footnote{Or, perhaps more precisely, expanding
  around the standard model (SM), since we are neither requiring nor
  precluding the existence of supersymmetry at the weak scale, though
we use the language of supersymmetry throughout.}
and exploring gauge sectors that may live nearby. In previous work
\cite{Cvetic:2011iq} we systematically added matter fields to MSSM
realizations that would otherwise violate string consistency
conditions; up to five matter fields beyond the exact MSSM spectrum
were added in the most general way possible in type II theories.
In this work we will further refine our study of this dataset, systematically
examining the possible sets of SM charges of exotics, the prevalence of singlets,
and their Yukawa couplings to other fields. Using
string consistency as a guiding principle, the most natural possibilities
for exotics are standard models singlets, vector-like
quarks and leptons, and triplets of $SU(2)_L$ that do not
carry hypercharge; these may be relevant for explaining the diphoton 
excess. Notably in all of these models, any vector-like exotics are
vector-like with respect to the standard model, but chiral with respect
to an additional $U(1)$, and hence it is plausible that they would
survive to low scale; in some scenarios their mass may be correlated
with the $\mu$-term or other scales in the visible sector, but we leave
such a study for future work. 

The second possibility we consider is a string embedding of a simplified
model. If the standard model is augmented by a pseudoscalar $\phi$ that
interacts with gluons and photons via the effective couplings
\begin{equation}
\cL_{S,eff} = \frac{c_{sBB}}{\Lambda_B} \,\, \phi \, F_Y \wedge F_Y + \frac{c_{sgg}}{\Lambda_G} \,\, \phi\, G \wedge G + \dots,
\end{equation}
where $F_Y$ and $G$ are the weak hypercharge and QCD field strength,
then there is a simple mechanism for $\phi$ production via gluon
interactions and a subsequent decay into two photons. Such axionic
couplings are ubiquitous in string theory, as they account for the
generalized Green-Schwarz mechanism of anomaly cancellation in string
theory compactifications. For example, in type II string theory they
arise from dimensional reduction of the of Wess-Zumino D-brane action.
(See \cite{Blumenhagen:2005mu} and references therein.)  We consider
these couplings in the context of type II compactifications in section
\ref{sec:axions}, focusing on the interplay between anomalous
$U(1)$'s, axionic couplings, and the low string scale necessary to
give decay rates into photons large enough to account for the diphoton
excess.

In section \ref{sec:excess} we provide a first look phenomenological
analysis of these models and in section \ref{sec:refined} we present
an extended phenomenological analysis that first appeared in our
February 2016 preprint addendum \cite{Cvetic:2016omj}.

\subsection{Rules for model building and exotics from string constraints}

Before beginning, we would like to summarize the rules for bottom-up
model building imposed by weakly coupled type II compactifications
with intersecting D-branes.  In appendix \ref{sec:D-brane appendix} we
give a more detailed overview of the basic ingredients of type II
orientifold compactifications and how they give rise to low energy
gauge sectors. These rules are important for the rest of the work, as
they are necessary for gauge sector embeddings into intersecting brane
models and they differ from those typically considered in quantum
field theory. We will state these rules concisely here; further details are
presented in the appendix.

The rules for model building in weakly coupled compactifications
with intersecting branes are somewhat rigid. We will use the language
of the type IIa theory, though the statements hold equally well for
type IIb or type I compactifications. The rules are:
\begin{itemize}
\item \textbf{\emph{Groups:}} $U(N)$ groups are obtained from wrapping $N$
  D6-branes on general cycles. Motivated by this, we consider groups of the
  form:
  \begin{equation}
    G = \prod_i U(N_i),
  \end{equation}
  though it would be interesting in future work to allow for an $Sp(1)$ factor,
  since this gives an alternative way of realizing $SU(2)_L$.
\item \textbf{\emph{Matter:}} May only be bifundamental under two
  $U(N)$ factors, or a symmetric or antisymmetric tensor of one
  factor. The orientifold image branes allow for the
  existence of all signs on bifundamentals;
  i.e. $(\fund_a,\ov\fund_b)$, $(\fund_a,\fund_b)$,
  $(\ov\fund_a,\fund_b)$, and $(\ov\fund_a,\ov\fund_b)$ are all
  possibilities, where $\fund_a$ ($\ov\fund_a$) denotes the
  fundamental (antifundamental) of $U(N_a)$.  Young tableaux is also
  used throughout for symmetric and antisymmetric tensor
  representations.
\item \textbf{\emph{Consistency:}} There are constraints on the chiral spectrum
  necessary for tadpole cancellation. They are:
  \begin{equation}
	\label{eqn:chiral tadpole constraint}
	T_a := \# a - \# \ov a + (N_a+4)\,\, (\# \, \Ysymm_a - \#\, \ov \Ysymm_a) + (N_a-4) \,\, (\# \, \Yasymm_a - \# \, \ov \Yasymm_a) = 0,
  \end{equation}
  where this is a modular constraint in the $N_a=1$ case due to the non-existence of antisymmetrics of $U(1)$; in
  other cases this constraint is an exact equality.

\item \textbf{\emph{Massless Hypercharge:}}  There are constraints on the chiral spectrum necessary for the
  absence of axionic couplings that would give a St\"uckelberg mass to $U(1)_Y$. For a $U(1)$ that is
  a linear combination $\sum q_i U(1)_i$ of the $U(1)_i\subset U(N_i)$ the constraint reads
\begin{equation}
\label{eqn:chiral masslessness constraint}
-q_aN_a\,\,(\#\Ysymm_a - \#\ov\Ysymm_a + \#\Yasymm_a - \#\ov\Yasymm_a) + \sum_{x\ne a} q_x N_x \,\, (\#(a,\ov x) - \#(a,x)) = 0
\end{equation}
for $N_a\ge 2$, and
\begin{equation}
\label{eqn:chiral masslessness constraint N1}
-q_a \,\,\frac{\#a - \#\ov a + 8(\#\Ysymm_a-\#\ov\Ysymm_a)}{3} + \sum_{x\ne a} q_x N_x \,\, (\#(a,\ov x) - \#(a,x)) = 0
\end{equation}
for $N_a=1$. The massless hypercharge constraint ensures that there is a linear combination satisfying these equations
that can be identified as hypercharge.
\item \textbf{\emph{Superpotential Couplings:}} In addition to their SM
  charges, all of the fields in the quivers\footnote{A quiver is a graphical representation of the gauge factors (nodes) and chiral matter (directed edges) of the theory.} we study will also carry
  charges under $U(1)$'s that are often anomalous, in which case
  the anomalies are cancelled via the generalized Green-Schwarz
  mechanism. These act as effective global symmetries on the low
  energy theory at the perturbative level. They impose selection rules on
  Yukawa couplings, as couplings that are SM-invariant may not be
  invariant with respect to the anomalous $U(1)$'s; throughout we will
  consider the holomorphic Yukawa couplings of the superpotential.
  However, D-brane instantons
  \cite{Blumenhagen:2006xt,Ibanez:2006da,Blumenhagen:2009qh} can
  generate these operators non-perturbatively, or the exotics could
  obtain masses via expectation values of singlets \cite{Cvetic:2010dz}.
\end{itemize}
In particular, throughout when we speak of a quiver, we mean one with
gauge groups and matter that follow these rules, and when we speak of
an MSSM quiver, we mean one whose chiral supermultiplets precisely
match the MSSM, without three right-handed neutrinos\footnote{The
  inclusion of three right-handed neutrino fields would complicate the
  analysis, but would most likely not change the results
  significantly.}.  When we speak of a consistent quiver, we mean one
with a chiral spectrum satisfying the conditions \eqref{eqn:chiral
  tadpole constraint} and also a massless hypercharge, even though the
latter is required only for phenomenological consistency rather than
theoretical consistency.

\vspace{.5cm} The conditions \eqref{eqn:chiral tadpole constraint}
necessary for tadpole cancellation play a critical role in this paper;
we will explain in section \ref{sec:exotics} how they can be
practically used to guide the addition of exotics. Here, though, we
would like to recall associated theoretical issues. Namely, for
$N_a>2$ the condition \eqref{eqn:chiral masslessness constraint} is
equivalent to the condition for the cancellation of cubic non-abelian
anomalies associated to $SU(N_a)$. However, tadpole cancellation also
requires that conditions be satisfied for $N_a=2,1$, and there are no
non-abelian anomalies for these groups: tadpole cancellation imposes
stronger constraints than typical anomaly cancellation
\cite{Uranga:2000xp,Aldazabal:2000dg}. The constraints
for $N_a=2,1$ are necessary but not sufficient for tadpole
cancellation, necessary but not sufficient for $U(1)_a\subset U(N_a)$
anomaly cancellation \cite{Ibanez:2001nd, Ghilencea:2002da}, and necessary \emph{and} sufficient
\cite{Halverson:2013ska} for $SU(M+N_a)^3$ anomaly cancellation if the
system were to nucleate $M$ brane anti-brane pairs, embedding
$U(N_a)$ into $U(M+N_a)$. To our knowledge, this anomaly nucleation is
the only known pathology that the constraints \eqref{eqn:chiral
  tadpole constraint} are necessary and sufficient for avoiding.  We
refer the reader to \cite{Halverson:2013ska} for an in depth
discussion of all of these issues.

\section{Scalars and Heavy Exotics}
\label{sec:exotics}

Given these model building rules, in this section we will recall how
our previous work \cite{Cvetic:2011iq} used the tadpole constraints to
expand around the MSSM, constructing all consistent quivers with up to
five exotics beyond the MSSM spectrum, adding them only for the sake
of consistency of the theory.  We will then turn to a new analysis,
studying in that dataset the SM quantum numbers of all possible exotic
sectors, those that involve singlets, those that involve one singlet
(and thus might present a simple model for the diphoton excess), and
also the perturbative couplings of models with one singlet.

\subsection{Results of a systematic analysis}

Let us review some of the basic results of \cite{Cvetic:2011iq}. We considered
three-stack and four-stack D-brane models, which have
\begin{equation}
G = U(3)_a\times U(2)_b \times U(1)_c \qquad \text{and} \qquad G = U(3)_a\times U(2)_b \times U(1)_c \times U(1)_d
\end{equation}
gauge symmetry, respectively. The hypercharge is a linear combination
\begin{equation}
Y = \sum q_i U(1)_i
\end{equation}
of the $U(1)$ factors, some of which are the trace parts of
$U(N)$'s. Such a linear combination is typically referred to as a
hypercharge embedding. It is possible to classify the possible
hypercharge embeddings for both three-stack and four-stack D-brane
models, and they are explicitly listed in section
\ref{sec:axions}. Once the hypercharge embedding is specified, it is
straightforward to classify all possible ways that each MSSM chiral
superfield may arise in the quiver, according to the rules discussed
in the introduction.

This gives all possible ways that the MSSM could potentially be
embedded for a particular hypercharge embedding and number of brane
stacks into intersecting D-brane configurations. However, most such
MSSM quivers do not satisfy the tadpole cancellation conditions
(\ref{eqn:chiral tadpole constraint}), in which case the embedding of
that MSSM sector into a D-brane compactification requires the addition
of exotics. This can be done systematically, given the rules for how
matter may arise, in order to make an inconsistent MSSM quiver into a
consistent quiver with the MSSM plus exotics.  This was the method
pursued\footnote{A similar approach where instead messengers to hidden
  sectors were added for consistency of the theory was pursued in
  \cite{Cvetic:2012kj}. Vector-like lepton dark matter motivated in
  part by such constraints was studied in \cite{Halverson:2014nwa} and
  singlet-extensions of the MSSM were also studied in
  \cite{Cvetic:2010dz}.} in \cite{Cvetic:2011iq}. Note in particular
that pairs of chiral multiplets that are vector-like with respect to
all symmetries of the theory never arise via this algorithm, since
they cannot contribute to (\ref{eqn:chiral tadpole constraint}) and
therefore they cannot make an inconsistent quiver consistent.
Furthermore, such completely non-chiral pairs would not be protected
from obtaining string scale masses.

\begin{table}[htb]
\centering
\scalebox{.95}{
\begin{tabular}{|c|c|c|c|c|}\hline
SM Rep & Total Multiplicity & Int. El. & $4^\text{th}$ Gen. Removed & Shifted $4^\text{th}$ Gen. Also Removed \\ \hline \hline
$(\textbf{1},\textbf{1})_{0}$ & $174276$ & $173578$ & $173578$ & $173578$ \\ \hline
$(\textbf{1},\textbf{3})_{0}$ & $48291$ & $48083$ & $48083$ & $48083$ \\ \hline
$(\textbf{1},\textbf{2})_{-\frac{1}{2}}$ & $39600$ & $39560$ & $38814$ & $38814$ \\ \hline
$(\textbf{1},\textbf{2})_{\frac{1}{2}}$ & $38854$ & $38814$ & $38814$ & $38814$ \\ \hline
$(\ov{\textbf{3}},\textbf{1})_{\frac{1}{3}}$ & $25029$ & $25007$ & $24261$ & $24241$ \\ \hline
$(\textbf{3},\textbf{1})_{-\frac{1}{3}}$ & $24299$ & $24277$ & $24277$ & $24241$ \\ \hline
$(\textbf{1},\textbf{1})_{1}$ & $15232$ & $15228$ & $14482$ & $14482$ \\ \hline
$(\textbf{1},\textbf{1})_{-1}$ & $14486$ & $14482$ & $14482$ & $14482$ \\ \hline
$(\ov{\textbf{3}},\textbf{1})_{-\frac{2}{3}}$ & $3501$ & $3501$ & $2755$ & $2755$ \\ \hline
$(\textbf{3},\textbf{1})_{\frac{2}{3}}$ & $2755$ & $2755$ & $2755$ & $2755$ \\ \hline
$(\textbf{3},\textbf{2})_{\frac{1}{6}}$ & $1784$ & $1784$ & $1038$ & $1038$ \\ \hline
$(\ov{\textbf{3}},\textbf{2})_{-\frac{1}{6}}$ & $1038$ & $1038$ & $1038$ & $1038$ \\ \hline
$(\textbf{1},\textbf{2})_{0}$ & $852$ & $0$ & $0$ & $0$ \\ \hline
$(\textbf{1},\textbf{2})_{\frac{3}{2}}$ & $220$ & $220$ & $220$ & $184$ \\ \hline
$(\textbf{1},\textbf{2})_{-\frac{3}{2}}$ & $204$ & $204$ & $204$ & $184$ \\ \hline
$(\textbf{1},\textbf{1})_{\frac{1}{2}}$ & $152$ & $0$ & $0$ & $0$ \\ \hline
$(\textbf{1},\textbf{1})_{-\frac{1}{2}}$ & $152$ & $0$ & $0$ & $0$ \\ \hline
$(\textbf{3},\textbf{1})_{\frac{1}{6}}$ & $124$ & $0$ & $0$ & $0$ \\ \hline
$(\ov{\textbf{3}},\textbf{1})_{-\frac{1}{6}}$ & $124$ & $0$ & $0$ & $0$ \\ \hline
$(\textbf{3},\textbf{1})_{-\frac{4}{3}}$ & $36$ & $36$ & $36$ & $0$ \\ \hline
$(\textbf{1},\textbf{3})_{-1}$ & $36$ & $36$ & $36$ & $0$ \\ \hline
$(\ov{\textbf{3}},\textbf{2})_{\frac{5}{6}}$ & $36$ & $36$ & $36$ & $0$ \\ \hline
$(\ov{\textbf{3}},\textbf{1})_{\frac{4}{3}}$ & $20$ & $20$ & $20$ & $0$ \\ \hline
$(\textbf{1},\textbf{3})_{1}$ & $20$ & $20$ & $20$ & $0$ \\ \hline
$(\textbf{3},\textbf{2})_{-\frac{5}{6}}$ & $20$ & $20$ & $20$ & $0$ \\ \hline
\end{tabular}}
\caption{Displayed are the standard model representation of matter additions obtained in \cite{Cvetic:2011iq}, together with their total multiplicity
  across all three-node and four-node quivers. The third column excludes quivers
  involving states that would lead to fractionally-charged color singlets. The fourth further excludes those where the matter additions correspond to a fourth generation, while the last also excludes a charge-shifted fourth generation. The remaining additions correspond to MSSM singlets, $SU(2)$ triplets with $Y=0$, and  quasichiral pairs.}\label{table:particle addition table}
\end{table}

The number of times exotic fields of a given SM charge occurred in the quivers of \cite{Cvetic:2011iq}
is reproduced in table \ref{table:particle addition table}. There are a few things we would
like to note about the results:
\begin{itemize}
\item String constraints have a clear preference for some multiplets over others!
\item Fourth families and charge-shifted fourth families are
  subtracted to arrive at the last column, along with multiplets
  giving rise to color-singlet states with fractional electric
  charge. From this column we see that these possibilities occurred a
  relatively small fraction of the time.
\item The most common exotic chiral multiplet is a standard model
  singlet $(\textbf{1},\textbf{1})_0$, and it occurs over three times
  as many times as any other exotic. Note that though it is a singlet
  with respect to the standard model, these singlets are always
  charged under anomalous $U(1)$'s, and therefore can have an
  intricate pattern of couplings to MSSM fields or to other exotics.
\item Next most frequent in the list is a chiral multiplet with SM charge $(\textbf{1},\textbf{3})_0$.
\item From there, we see a variety of exotic quarks and leptons, and
  from the relative multiplicities in the last column it is easy to
  see that they often come as pairs that are vector-like
  with respect to the SM.
\item Any such pairs are vector-like with respect to the standard model, but chiral
  with respect to some $U(1)$, which is often anomalous. We call such vector pairs quasichiral.
\end{itemize}

These preferences and statistics of quivers represent the
  possible (small) extensions of the MSSM that may be embedded in type
  II intersecting D-brane models, which may or may not be correlated with statistics
  of global string embeddings that realize them or of cosmological
  vacuum selection. 

\subsection{Models with singlets and charged exotics}

In this section we would like to explore further aspects of the
dataset of \cite{Cvetic:2011iq} as it relates to the possible
\emph{sets} of exotics, those sets with multiple singlets, those sets
with one singlet, and then we will discuss perturbative couplings of
the singlets in models with one singlet, the structure of which (at
the perturbative level) is dictated by the anomalous $U(1)$ charges of
the chiral multiplets.

In table \ref{table:all Exotics binned} we present all possible
exotics that arise in the quivers, binned according to their SM
charges. In each case, the number of quivers with that set of exotics
is listed. Within the quivers for a given set of exotics the anomalous
$U(1)$ charges of fields may differ, and therefore so does the
perturbative superpotential. We will return to a study of singlet
couplings below. Just as table \ref{table:particle
  addition table} demonstrates that some exotics are much preferred over
others, table \ref{table:all Exotics binned} demonstrates that some exotic \emph{sets}
are much preferred. The six most common sets are
\begin{center}
$\{(\textbf{1},\textbf{1})_{0}$ $(\textbf{1},\textbf{1})_{0}$ $(\textbf{1},\textbf{2})_{-\frac{1}{2}}$ $(\textbf{1},\textbf{2})_{\frac{1}{2}}$ $(\textbf{1},\textbf{1})_{0}\}$, \\
$\{(\textbf{1},\textbf{3})_{0}$ $(\textbf{1},\textbf{1})_{0}$ $(\textbf{1},\textbf{1})_{0}$ $(\textbf{1},\textbf{2})_{-\frac{1}{2}}$ $(\textbf{1},\textbf{2})_{\frac{1}{2}}\}$, \\
$\{(\textbf{1},\textbf{1})_{0}$ $(\textbf{1},\textbf{1})_{0}$ $(\textbf{1},\textbf{1})_{0}$ $(\textbf{1},\textbf{1})_{0}$ $(\textbf{1},\textbf{1})_{0}\}$, \\
$(\textbf{1},\textbf{1})_{0}$ $(\textbf{1},\textbf{1})_{0}$ $(\textbf{1},\textbf{1})_{0}$ $(\textbf{3},\textbf{1})_{-\frac{1}{3}}$ $(\ov{\textbf{3}},\textbf{1})_{\frac{1}{3}}\}$, \\
$\{(\textbf{1},\textbf{3})_{0}$ $(\textbf{1},\textbf{1})_{0}$ $(\textbf{1},\textbf{1})_{0}$ $(\textbf{1},\textbf{1})_{0}$ $(\textbf{1},\textbf{1})_{0}\}$,\\
$\{(\textbf{1},\textbf{2})_{\frac{1}{2}}$ $(\textbf{1},\textbf{2})_{-\frac{1}{2}}$ $(\textbf{1},\textbf{1})_{0}$ $(\textbf{1},\textbf{1})_{0}\}$.
\end{center}
We see an interesting array of exotic sectors. All have more than one singlet, and there are often vector-like quarks, vector-like
leptons, or the $Y=0$ $SU(2)_L$ triplet. Less prevalent, though still very common, are models with exotic sectors of
the form
\begin{center}
 $(\textbf{1},\textbf{2})_{\frac{1}{2}}$ $(\textbf{1},\textbf{2})_{-\frac{1}{2}}$ $(\textbf{1},\textbf{1})_{0}$ $(\ov{\textbf{3}},\textbf{1})_{\frac{1}{3}}$ $(\textbf{3},\textbf{1})_{-\frac{1}{3}}$,
\end{center}
which have a singlet, as well as exotic vector-pairs of both down-type quarks and leptons. There are 
many other interesting possibilities in table \ref{table:all Exotics binned}, some of which also occur frequently, though some exotic sets occur very
infrequently.

In tables \ref{table:Exotics with S binned} and \ref{table:Exotics with one S binned} we present the same
data in a different form. In the former we restrict to exotic sets with at least one singlet, and in the
latter we restrict to cases with exactly one singlet, since these two restrictions are perhaps most relevant
for explaining the diphoton excess.

Finally, we have also performed an analysis of all renormalizable,
perturbative couplings of singlets, including in particular their
Yukawa couplings. The presence or absence of these couplings is
determined by the SM charges and anomalous $U(1)$ charges of fields in
the quiver, including the exotics. The results are presented in tables
\ref{table:One S couplings 1}-\ref{table:One S couplings 9} in the
appendix. The tables should be read as follows. In the first column is
a set of exotics with their SM charges, and the second contains its
multiplicity; suppose it is $N$. These $N$ quivers differ in their
anomalous $U(1)$ charges, which can affect the allowed couplings of
singlets to other fields. Therefore, given a set of exotics in column
one we list all the possible singlet couplings to fields of certain SM
charges in column three, and the associated number of quivers with
those singlet couplings in column four.

One important point about the quivers that is captured in the
coupling tables is worth discussing.  Fix a quiver, and suppose
that the exotic set consisted of a singlet $S$ and vector-like leptons
$X$ and $\ov X$ with hypercharge $\pm\frac12$ respectively. Depending
on its couplings to other fields in the quiver, it may be that $\ov X$
is better identified as $H_d$ or $L$, and one of those is better
identified as an exotic; that is, the labels are arbitrary and
applying them correctly depends on the couplings. Therefore, in such a
quiver one should really consider the perturbative singlet couplings
to \emph{all} fields, since this ambiguity may arise. For example,
collecting some fields with the same SM charge
\begin{equation}
F_i \in \{ L_1, L_2, L_3, H_d, \ov X\}, \qquad \ov F_j \in \{H_u, X\},
\end{equation}
all perturbatively allowed singlet couplings in
\begin{equation}
W = \lambda_{ij} \,\, S\, F_i \ov F_j + \dots
\end{equation}
should be considered. In this model, there would be $10$
\emph{possible} couplings of the form $SF\ov F$ according to the
$2\times 5$ matrix, but we compute how many are there at the
perturbative level by determining invariance under anomalous $U(1)$'s.
If only the couplings involving $X$ were allowed, but not $H_u$, then
we write
$5\times ((\textbf{1},\textbf{1})_0 (\textbf{1},\textbf{2})_{-\frac12}
(\textbf{1},\textbf{2})_{\frac12})$
in the table, denoting that there are $5$ perturbatively allowed
couplings. Similar comments apply to the other fields.

Examining the coupling tables in the appendix, one sees that there
are a broad variety of possible singlet couplings to MSSM fields or
other exotics in these quivers. Often the couplings are to both types
of fields, but typically after diagonalization of the mass matrices
they only have significant couplings to the massive exotic pair. We 
study such models in detail in section \ref{sec:excess}.

\begin{table}[htb]
\begin{center}
\scalebox{.55}{
\centering
\begin{tabular}{|c|c|c|c|c|c|} \hline
Exotics & Multiplicity & Exotics & Multiplicity & Exotics & Multiplicity \\ \hline
$(\textbf{1},\textbf{1})_{0}$ $(\textbf{1},\textbf{1})_{0}$ $(\textbf{1},\textbf{2})_{-\frac{1}{2}}$ $(\textbf{1},\textbf{2})_{\frac{1}{2}}$ $(\textbf{1},\textbf{1})_{0}$  & $5232$ & $(\textbf{1},\textbf{3})_{0}$ $(\textbf{1},\textbf{1})_{0}$ $(\textbf{1},\textbf{1})_{0}$ $(\textbf{1},\textbf{2})_{-\frac{1}{2}}$ $(\textbf{1},\textbf{2})_{\frac{1}{2}}$  & $5018$ & $(\textbf{1},\textbf{1})_{0}$ $(\textbf{1},\textbf{1})_{0}$ $(\textbf{1},\textbf{1})_{0}$ $(\textbf{1},\textbf{1})_{0}$ $(\textbf{1},\textbf{1})_{0}$  & $4230$ \\
$(\textbf{1},\textbf{1})_{0}$ $(\textbf{1},\textbf{1})_{0}$ $(\textbf{1},\textbf{1})_{0}$ $(\textbf{3},\textbf{1})_{-\frac{1}{3}}$ $(\ov{\textbf{3}},\textbf{1})_{\frac{1}{3}}$  & $3893$ & $(\textbf{1},\textbf{3})_{0}$ $(\textbf{1},\textbf{1})_{0}$ $(\textbf{1},\textbf{1})_{0}$ $(\textbf{1},\textbf{1})_{0}$ $(\textbf{1},\textbf{1})_{0}$  & $3646$ & $(\textbf{1},\textbf{2})_{\frac{1}{2}}$ $(\textbf{1},\textbf{2})_{-\frac{1}{2}}$ $(\textbf{1},\textbf{1})_{0}$ $(\textbf{1},\textbf{1})_{0}$  & $3618$ \\
$(\textbf{1},\textbf{1})_{0}$ $(\textbf{1},\textbf{2})_{-\frac{1}{2}}$ $(\textbf{1},\textbf{2})_{-\frac{1}{2}}$ $(\textbf{1},\textbf{2})_{\frac{1}{2}}$ $(\textbf{1},\textbf{2})_{\frac{1}{2}}$  & $3340$ & $(\ov{\textbf{3}},\textbf{1})_{\frac{1}{3}}$ $(\textbf{1},\textbf{3})_{0}$ $(\textbf{1},\textbf{1})_{0}$ $(\textbf{1},\textbf{1})_{0}$ $(\textbf{3},\textbf{1})_{-\frac{1}{3}}$  & $3257$ & $(\textbf{1},\textbf{1})_{0}$ $(\textbf{1},\textbf{1})_{0}$ $(\textbf{1},\textbf{1})_{0}$ $(\textbf{1},\textbf{1})_{0}$  & $3106$ \\
\hline
\end{tabular}}
\end{center}
\caption{The multiplicities of exotic sectors, binned according to their SM quantum numbers. The multiplicities decrease left to right, then down rows.}
\label{table:all Exotics binned}
\end{table}

\begin{table}[htb]
\begin{center}
\scalebox{.55}{
\centering
\begin{tabular}{|c|c|c|c|c|c|} \hline
Exotics & Multiplicity & Exotics & Multiplicity & Exotics & Multiplicity \\ \hline
$(\textbf{1},\textbf{1})_{0}$ $(\textbf{1},\textbf{1})_{0}$ $(\textbf{1},\textbf{2})_{-\frac{1}{2}}$ $(\textbf{1},\textbf{2})_{\frac{1}{2}}$ $(\textbf{1},\textbf{1})_{0}$  & $5232$ & $(\textbf{1},\textbf{3})_{0}$ $(\textbf{1},\textbf{1})_{0}$ $(\textbf{1},\textbf{1})_{0}$ $(\textbf{1},\textbf{2})_{-\frac{1}{2}}$ $(\textbf{1},\textbf{2})_{\frac{1}{2}}$  & $5018$ & $(\textbf{1},\textbf{1})_{0}$ $(\textbf{1},\textbf{1})_{0}$ $(\textbf{1},\textbf{1})_{0}$ $(\textbf{1},\textbf{1})_{0}$ $(\textbf{1},\textbf{1})_{0}$  & $4230$ \\
$(\textbf{1},\textbf{1})_{0}$ $(\textbf{1},\textbf{1})_{0}$ $(\textbf{1},\textbf{1})_{0}$ $(\textbf{3},\textbf{1})_{-\frac{1}{3}}$ $(\ov{\textbf{3}},\textbf{1})_{\frac{1}{3}}$  & $3893$ & $(\textbf{1},\textbf{3})_{0}$ $(\textbf{1},\textbf{1})_{0}$ $(\textbf{1},\textbf{1})_{0}$ $(\textbf{1},\textbf{1})_{0}$ $(\textbf{1},\textbf{1})_{0}$  & $3646$ & $(\textbf{1},\textbf{2})_{\frac{1}{2}}$ $(\textbf{1},\textbf{2})_{-\frac{1}{2}}$ $(\textbf{1},\textbf{1})_{0}$ $(\textbf{1},\textbf{1})_{0}$  & $3618$ \\
$(\textbf{1},\textbf{1})_{0}$ $(\textbf{1},\textbf{2})_{-\frac{1}{2}}$ $(\textbf{1},\textbf{2})_{-\frac{1}{2}}$ $(\textbf{1},\textbf{2})_{\frac{1}{2}}$ $(\textbf{1},\textbf{2})_{\frac{1}{2}}$  & $3340$ & $(\ov{\textbf{3}},\textbf{1})_{\frac{1}{3}}$ $(\textbf{1},\textbf{3})_{0}$ $(\textbf{1},\textbf{1})_{0}$ $(\textbf{1},\textbf{1})_{0}$ $(\textbf{3},\textbf{1})_{-\frac{1}{3}}$  & $3257$ & $(\textbf{1},\textbf{1})_{0}$ $(\textbf{1},\textbf{1})_{0}$ $(\textbf{1},\textbf{1})_{0}$ $(\textbf{1},\textbf{1})_{0}$  & $3106$ \\
\hline
\end{tabular}}
\end{center}
\caption{The multiplicities
of exotic sectors that contain singlets.}
\label{table:Exotics with S binned}
\end{table}

\begin{table}[htb]
\begin{center}
\scalebox{.55}{
\centering
\begin{tabular}{|c|c|c|c|c|c|} \hline
Exotics & Multiplicity & Exotics & Multiplicity & Exotics & Multiplicity \\ \hline
$(\textbf{1},\textbf{1})_{0}$ $(\textbf{1},\textbf{2})_{-\frac{1}{2}}$ $(\textbf{1},\textbf{2})_{-\frac{1}{2}}$ $(\textbf{1},\textbf{2})_{\frac{1}{2}}$ $(\textbf{1},\textbf{2})_{\frac{1}{2}}$  & $3340$ & $(\textbf{1},\textbf{2})_{\frac{1}{2}}$ $(\textbf{1},\textbf{2})_{-\frac{1}{2}}$ $(\textbf{1},\textbf{1})_{0}$ $(\ov{\textbf{3}},\textbf{1})_{\frac{1}{3}}$ $(\textbf{3},\textbf{1})_{-\frac{1}{3}}$  & $2429$ & $(\textbf{1},\textbf{3})_{0}$ $(\textbf{1},\textbf{1})_{0}$ $(\textbf{1},\textbf{2})_{-\frac{1}{2}}$ $(\textbf{1},\textbf{2})_{\frac{1}{2}}$  & $2354$ \\
$(\ov{\textbf{3}},\textbf{1})_{\frac{1}{3}}$ $(\textbf{1},\textbf{3})_{0}$ $(\textbf{1},\textbf{1})_{0}$ $(\textbf{3},\textbf{1})_{-\frac{1}{3}}$  & $1744$ & $(\textbf{1},\textbf{2})_{\frac{1}{2}}$ $(\textbf{1},\textbf{2})_{-\frac{1}{2}}$ $(\textbf{1},\textbf{1})_{0}$  & $1728$ & $(\textbf{1},\textbf{1})_{1}$ $(\textbf{1},\textbf{1})_{-1}$ $(\textbf{1},\textbf{2})_{-\frac{1}{2}}$ $(\textbf{1},\textbf{1})_{0}$ $(\textbf{1},\textbf{2})_{\frac{1}{2}}$  & $1496$ \\
$(\textbf{1},\textbf{1})_{0}$ $(\ov{\textbf{3}},\textbf{1})_{\frac{1}{3}}$ $(\ov{\textbf{3}},\textbf{1})_{\frac{1}{3}}$ $(\textbf{3},\textbf{1})_{-\frac{1}{3}}$ $(\textbf{3},\textbf{1})_{-\frac{1}{3}}$  & $1462$ & $(\textbf{1},\textbf{3})_{0}$ $(\textbf{1},\textbf{3})_{0}$ $(\textbf{1},\textbf{1})_{0}$ $(\textbf{1},\textbf{2})_{-\frac{1}{2}}$ $(\textbf{1},\textbf{2})_{\frac{1}{2}}$  & $1354$ & $(\textbf{1},\textbf{3})_{0}$ $(\textbf{1},\textbf{1})_{0}$  & $1168$ \\
\hline
\end{tabular}}
\end{center}
\caption{The multiplicities of exotic sectors that have exactly one singlet. }
\label{table:Exotics with one S binned}
\end{table}

\subsection{Generalized analysis and remarks}

Though we derived many concrete results in the previous section, there
are results that are quite general in this class of models that can be
understood analytically. In this section we would like to discuss a
few of them.

First, we emphasize that all vector-like
exotics $X, \ov X$ that appear in these models are chiral with respect to one or
more $U(1)$'s, and therefore the associated mass term is forbidden in
perturbation theory. However, 
the masses may
be generated by a singlet VEV
\begin{equation}
W_{eff} = \lambda_M \langle S_M\rangle \, X \ov X + \dots,
\end{equation}
where in many cases $S_M$ may be the field associated with the
diphoton excess\footnote{Since the anomalous $U(1)$ symmetry is
  effectively a global symmetry, both scalar and pseudoscalar
  components will typically survive.}. They may also may be generated
non-perturbatively in some cases by D-brane instantons
\cite{Blumenhagen:2006xt,Ibanez:2006da,Blumenhagen:2009qh} that break
the anomalous $U(1)$ to a discrete subgroup. In this case they have
the form
\begin{equation}
W_{\text{non-pert}} = e^{-T}\, X \ov X + \dots,
\end{equation}
where $T$ is a chiral superfield that completes a metric
modulus of the compactification. 

These vector-pairs are chiral with respect to a non-SM symmetry (i.e.,
quasichiral) and are therefore protected from receiving a string scale
mass. Depending on details of the compactification, and possible
correlations with other couplings (for example, if the same effect
that generates the $\mu$-term also generates a mass for $X\ov X$),
these fields may be present at the electroweak scale.

Second, consider an operator $\cO$ and its possible coupling to $S$,
namely $S\cO$. For singlets typically considered in the literature,
and in particular considered in many works on the diphoton excess, it
is assumed that if $\cO$ is an operator allowed by the symmetries,
then so is $S\cO$; after all, $S$ is a singlet. In contrast, in all
of the models of this section $S$ is a particular type of field
arising from an open string ending on two intersecting D-branes;
though it is a SM singlet, it is charged under one or more anomalous
$U(1)$'s. Thus, if the perturbative
superpotential contains
\begin{equation}
W_{\text{pert}} = \lambda \cO + \cdots ,
\end{equation}
then $S\cO$ \emph{never} appears in the perturbative
superpotential. This provides a certain type of barrier to the visible
sector; for example, for any perturbatively allowed MSSM Yukawa
coupling $\cO_Y$, $S\cO_Y$ is forbidden by anomalous $U(1)$'s. This is
perhaps most relevant if the top-quark Yukawa couplings is $O(1)$ due
to being perturbatively allowed; in this case
the interactions of $S$ with the top quark are reduced, and in
particular $S Q_L H_u u_l^c$ is either completely absent,
non-perturbative, or otherwise suppressed.

Finally, the field $T\sim (\textbf{1},\textbf{3})_0$ occurs very
frequently in these quivers but has particular couplings. It
necessarily arises as a symmetric tensor of $U(2)_b$, where the
$SU(2) \subset U(2)_b$ is $SU(2)_L$. Therefore it has
charge $\pm 2$ under $U(1)_b\subset U(2)_b$, and this forbids all
singlet couplings of the form $STT$. Thus, though many quivers contain
$T$, there is never a perturbative Yukawa coupling that allows for $S$
decay into $TT$; only some of the exotics can run in the loops that
contribute to singlet decay into photons.

\section{Pseudoscalars and Hypercharge Embedding}
\label{sec:axions}

Within string compactifications there is another natural mechanism for
production and decay of pseudoscalars (axions), related to the
generalized anomaly cancellation mechanism.

While consistent string constructions of course have to be compatible
with standard field theory anomaly cancellations for the non-abelian
gauge fields as well as the non-anomalous $U(1)$'s, such as the
hypercharge, string theory provides further constraints due to the
Green-Schwarz mechanism: these are constraints that arise due to
triangular anomaly cancellation for anomalous $U(1)$'s via
exchanges of the string axions coupled to the Chern-Simons terms.

As a prototype, gauge theories in type II string compactifications  not only contain non-anomalous nonabelian and abelian factors, as in the standard model, but also generically include anomalous $U(1)$ factors of the trace generator of a $U(N)$ factor.  Matter fields of the same standard model representation can  in principle carry different charges with respect to the anomalous $U(1)$'s,  as they correspond to the appearance at the intersection of different pairs of D-branes.
The Wess-Zumino component of the D-brane worldvolume action provides the necessary Chern-Simons couplings, responsible for the cancellation of abelian and mixed anomalies associated with the anomalous $U(1)$'s.

The structure of four-dimensional Chern-Simons terms is of the form:
 \begin{equation}
\frac{\phi_i}{M_{st}}  Tr( F_i \wedge F_i) \, , \quad \frac{\phi_i}{M_{st}}  R \wedge R\, , \quad    M_{st}\, B_i\wedge Tr(F_i)\, , \label{CS}
 \end{equation}
 where $F_i$ are the field strengths of the $U(N)_i$ gauge bosons and
 a specific combination of $0$-forms $\phi_i$ and its Hodge dual
 $2$-forms $B_i$ both possess an axionic shift symmetry.  (For details
 and references therein, see, e.g.,
 \cite{Blumenhagen:2005mu,Cvetic:2011vz}.)  The coefficients of the
 linear combinations of the $0$- and $2$-forms appear precisely in a
 combination that cancel all abelian and mixed gauge anomalies.
 Furthermore the last term above is also responsible for a generic
 appearance of the St{\"u}ckelberg mass for the anomalous
 $U(1)_i$.\footnote{These terms can in principle be included directly
   in the study of effective theory with the same conclusions. See,
   \cite{Anastasopoulos:2006cz,Mambrini:2009ad}.}  Furthermore the
 term $\phi_i R \wedge R$ is necessary for cancellation of mixed
 abelian-gravitational anomalies \cite{Cvetic:2001nr}.

It is the Chern-Simons coupling of an non-anomalous $U(1)_Y$ that can
be responsible for the decay channel of a string axion into diphotons:
\begin{equation}
\frac{c_Y}{M_{st}} \,   \phi Tr( F_Y \wedge F_Y) \,  , \label{CSY}
 \end{equation}
 where $F_Y$ are the hypercharge boson field strength and $\phi$ is the
 (normalized) axion field.  Since $U(1)_Y$ is non-anomalous the axion
 Hodge dual $2$-form field $B$ should not have a coupling to the $U(1)_Y$
 field strength, i.e.  $B_Y\wedge Tr(F_Y)$ , and thus should not have
 a St{\"u}ckelberg mass.

 In the three stack Standard Model constructions with
 $U(3)_a\times U(2)_b\times U(1)_c$ this results in the two choices
 for $U(1)_Y$: \cite{Anastasopoulos:2006da}:
\begin{equation}
U(1)_Y=\frac{1}{6}U(1)_a+\frac{1}{2}U(1)_c\, , \qquad
U(1)_Y=-\frac{1}{3}U(1)_a-\frac{1}{2}U(1)_b\, . 
\end{equation}
In the four stack Standard Model constructions with
$U(3)_a\times U(2)_b\times U(1)_c\times U(1)_d$ there are six
possibilities for the hypercharge \cite{Anastasopoulos:2006da}, many of which were
already determined in \cite{Gmeiner:2005vz}:
\begin{align}
\label{eqn:4-node hypercharge embeddings}
U(1)_Y &= -\frac{1}{3} U(1)_a - \frac{1}{2}U(1)_b\, ,  \qquad\qquad \qquad \,\,\,\,\,\,\,\,\,\,
U(1)_Y = -\frac{1}{3} U(1)_a - \frac{1}{2}U(1)_b + \frac{1}{2}U(1)_d\, , \notag \\
U(1)_Y &= -\frac{1}{3} U(1)_a - \frac{1}{2}U(1)_b + U(1)_d\, ,  \qquad\qquad
U(1)_Y = \frac{1}{6} U(1)_a + \frac{1}{2}U(1)_c\, ,   \qquad \qquad \qquad\notag \\
U(1)_Y &= \frac{1}{6} U(1)_a + \frac{1}{2}U(1)_c + \frac{1}{2}U(1)_d\, ,  \qquad \qquad
U(1)_Y = \frac{1}{6} U(1)_a + \frac{1}{2}U(1)_c + \frac{3}{2}U(1)_d\, .
\end{align}
In all these cases the component of $U(1)_a$  in $U(1)_Y$  is either $\frac{1}{3}$ or $-\frac{1}{6}$. 

Thus, the D-brane compactifications naturally provide a decay diphoton
channel via Chern-Simons coupling (\ref{CSY}) to hypercharge gauge
bosons.  Note, however, for the couplings (\ref{CSY}) to produce a large
enough signal, the string scale $M_{st}$ has to be low, i.e. in the
TeV regime.

A possible production channel could in principle be due to the
Chern-Simons terms
\begin{equation}
\frac{c_3}{M_{st}}  \phi \ Tr( G \wedge G) \,  , \qquad
 \end{equation}
 where $G$ corresponds to the gluon field strength and $\phi$ is a
 (normalized) axion field that also couples to (\ref{CSY}).  However,
 $\phi$'s $2$-form Hodge dual should not couple to $U(1)_a$
 to avoid a $U(1)_a$ St{\"u}ckelberg mass due to $\phi$ and thus resulting
 in a $U(1)_a$ gauge boson mass at 750 GeV.

We however note that the weakly coupled D-brane constructions
necessarily result in $U(1)_a\times SU(2)_b^2$ mixed anomalies due to
the presence of $Q_L$'s, which cannot be cancelled, unless one
introduces ``mirror'' quark doublets. Therefore, the St{\"u}ckelberg term $B_a\wedge Tr(F_a)$ has to be present. However, the $2$-form $B_a$ field can be Hodge dual to a different axion field
$\phi_a$. Note that the Ramond-Ramond axion fields are ubiquitous in Type II compactifications and thus a different $\phi_a$ can participates in $U(1)_a$ anomaly cancellation. Concrete constructions of this type would require further analysis.

\section{On the $750$ GeV Diphoton Excess: A First Look}
\label{sec:excess}

In this section we examine whether the singlets and exotics most
frequently found in our quivers of section \ref{sec:exotics} may
account for the diphoton excess, reserving a more refined analysis for
the next section. We leave a detailed analysis of the low string scale
scenario of section \ref{sec:axions} to future work.  A number of
works have already appeared on explaining the diphoton rate from
decays of singlets into loops of charged exotics. In this section we
will not try to address all of those works, but will instead base our
analysis on Franceschini et al. in \cite{Franceschini:2015kwy}.

The perturbative superpotential couplings of singlets to other exotics
discussed in section \ref{sec:exotics} induce Yukawa couplings of a
complex scalar boson $s$ of mass $M$ to exotic Dirac fermions $X$ with
mass $M_i$, charge $Q_i$, and
color representation $r_i$ of the form
\begin{equation}
s \ov X(y_i + i y_{5,i}\gamma_5) X,
\end{equation}
where the scalar versus pseudoscalar interaction depends on the detailed
properties of $s$. The decay widths into $GG$ and $\gamma\gamma$
induced by associated fermion loops are \cite{Djouadi:2005gi,Franceschini:2015kwy}
\begin{align}
\Gamma(s\to GG) &= M \frac{\alpha_3^2}{2\pi^3} \left| \sum_i T^3(i) \sqrt{\tau_i} y_i S(\tau_i) \right|^2 \nonumber \\
\Gamma(s\to \gamma\gamma) &= M\frac{\alpha^2}{16\pi^3} \left|\sum_i n_3(i) Q_i^2 \sqrt{\tau_i} y_i S(\tau_i) \right|^2
\end{align}
for $s$ a scalar, and 
\begin{align}
\label{eq:pseudo rates}
\Gamma(s\to GG) &= M \frac{\alpha_3^2}{2\pi^3} \left| \sum_i T^3(i) \sqrt{\tau_i} y_{5i}P(\tau_i) \right|^2 \nonumber \\
\Gamma(s\to \gamma\gamma) &= M\frac{\alpha^2}{16\pi^3} \left|\sum_i n_3(i) Q_i^2 \sqrt{\tau_i} y_{5i}P(\tau_i)\right|^2
\end{align}
for $s$ a pseudoscalar, where $T^3$ and $n_3$ are the Dynkin index and
dimension of the color representation $r_i$ and
\begin{equation}
P(\tau) = \arctan^2(1/\sqrt{\tau-1}), \qquad S(\tau) = 1 + (1-\tau)P(\tau),
\end{equation}
with $\tau_i = 4M_i^2/M^2$ and $M$ the mass of $s$. In the scalar case
additional loops involving spin-0 particles are possible, with
couplings given by $A$-terms in the supersymmetric case.

We first orient ourselves by making some optimistic assumptions to maximize the decay
rates $\Gamma(s \to GG)$ and $\Gamma(s \to \gamma\gamma)$. We assume a pseudoscalar decay since it gives a larger rate, take
each $M_i = M/2$, and assume that each Yukawa couplings $y_{5i}=1$.
Given these assumptions, the rates \eqref{eq:pseudo rates} simplify
to 
\begin{align}
\label{eq:rates our assumptions}
\frac{\Gamma(s \to GG)}{M} \simeq 9.82 \times 10^{-4} \left|\sum_i C_{GG}\right|^2  \nonumber \\
\frac{\Gamma(s \to \gamma\gamma)}{M} \simeq 7.49 \times 10^{-7} \left|\sum_i C_{\gamma\gamma}\right|^2,
\end{align}
where the only information left in the sum is the group theoretic
data.  Studying singlet couplings to exotics that arise in our
quivers, as listed in appendix \ref{appendix:tables}, we compute the
decay rates of the singlet in table \ref{table:decay table}. Comparing
to the left side of figure 1 of \cite{Franceschini:2015kwy}, we see
that the last five entries of our table \ref{table:decay table} may
account for a narrow diphoton resonance at $750$ GeV; in particular,
it falls on their blue band where $\Gamma_{tot} = \Gamma_{GG} +
\Gamma_{\gamma\gamma}$.  Vector-like down-type quarks by themselves,
on the other hand, do not have a high enough decay rate into photons
to account for the signal. The width could be larger if one allows for
the possibility $M_i < M/2$, or for decays into exotics without color
and/or electric charges, but then one would need additional exotics to
increase $\Gamma(s \to GG)$ and $\Gamma(s \to \gamma\gamma)$.

We would like to emphasize that most of the possible singlet couplings
in appendix \ref{appendix:tables} give rise to one of the scenarios in
table \ref{table:decay table},
sometimes by a field redefinition so that scenarios which look like they
have many fields of a given SM charge coupling to $s$ in fact have one
after the field redefinition.

We also emphasize that this analysis is not the result of
phenomenological model building, but instead utilizes string consistency
conditions to
necessitate (in most cases) the addition of exotics, which almost always
include singlets with interesting couplings to other fields. In many cases these singlets can couple to heavy
fermion exotics and account for the excess in ATLAS and CMS diphoton
searches at $750$ GeV.

\begin{table}[htb]
\centering
\scalebox{1.}{
\begin{tabular}{|c|c|c|c|c|}\hline
Representations& 
$C_{\gamma\gamma}$ &$ \Gamma_{\gamma\gamma}/M$& $C_{GG}$ &$ \Gamma_{GG}/M$  \\ \hline \hline
$(\textbf{1},\textbf{2})_{\frac{1}{2}}$ & 1 & $7.5 \times 10^{-7}$ & 0 & 0 \\ \hline
$2 \times (\textbf{1},\textbf{2})_{\frac{1}{2}}$ & 2 & $3.0 \times 10^{-6}$ & 0 & 0 \\ \hline
$ (\textbf{1},\textbf{1})_{1}$ &  1 & $7.5 \times 10^{-7}$ & 0 & 0 \\ \hline
$2 \times (\textbf{1},\textbf{1})_{1}$ & 2 & $3.0 \times 10^{-6}$ & 0 & 0 \\ \hline
$(\textbf{1},\textbf{1})_{1} +(\textbf{1},\textbf{2})_{\frac{1}{2}} $ & 2 & $3.0 \times 10^{-6}$ & 0 & 0 \\ \hline
$(\textbf{1},\textbf{2})_{\frac{3}{2}}$ & 5 & $1.9 \times 10^{-5}$ & 0 & 0 \\ \hline
$(\textbf{3},\textbf{1})_{-\frac{1}{3}}$ & $1/3$ & $8.3\times 10^{-8}$ & $1/2$ & $2.5 \times 10^{-4}$  \\ \hline
$2 \times (\textbf{3},\textbf{1})_{-\frac{1}{3}}$ & $2/3$ & $3.3\times 10^{-7}$ & $1$ & $9.8 \times 10^{-4}$  \\ \hline
$(\textbf{3},\textbf{1})_{-\frac{1}{3}} +(\textbf{1},\textbf{2})_{\frac{1}{2}} $ & $4/3$ & $1.3\times 10^{-6}$ & $1/2$ & $2.5 \times 10^{-4}$  \\ \hline
$(\textbf{3},\textbf{1})_{-\frac{1}{3}} +(\textbf{1},\textbf{1})_{1} $ & $4/3$ & $1.3\times 10^{-6}$ & $1/2$ & $2.5 \times 10^{-4}$  \\ \hline
$(\textbf{3},\textbf{2})_{\frac{1}{6}}$ & $5/3$ & $2.1\times 10^{-6}$ & $1$ & $9.8 \times 10^{-4}$  \\ \hline
$(\textbf{3},\textbf{1})_{\frac{2}{3}}$ & $4/3$ & $1.3\times 10^{-6}$ & $1/2$ & $2.5 \times 10^{-4}$  \\ \hline
$(\textbf{3},\textbf{1})_{-\frac{1}{3}} +(\textbf{3},\textbf{1})_{\frac{2}{3}}$ &$5/3$ & $2.1\times 10^{-6}$ & $1$ & $9.8 \times 10^{-4}$  \\ \hline
\end{tabular}}
\caption{Exotic sets with perturbative couplings to the singlet pseudoscalar. In each case the conjugate is included. The widths assume $\alpha=1/128$, $\alpha_s=0.1$, $y=1$, and $M_i=M/2=375$ GeV. The widths would be reduced by a factor $\sim$6.1  for a singlet scalar.}
\label{table:decay table}
\end{table}

\section{A Refined Phenomenological Analysis}
\label{sec:refined}
In this section\footnote{This analysis was first presented in our
  preprint addendum \cite{Cvetic:2016omj} in February 2016. Up to the
  appearance of that work, many other works
  \cite{Franceschini:2015kwy,McDermott:2015sck,Ellis:2015oso,Gupta:2015zzs,Martinez:2015kmn,Fichet:2015vvy,Bian:2015kjt,Falkowski:2015swt,Bai:2015nbs,Dhuria:2015ufo,Chakraborty:2015jvs,Wang:2015kuj,Murphy:2015kag,Hernandez:2015ywg,Huang:2015rkj,Badziak:2015zez,Cvetic:2015vit,Cheung:2015cug,Zhang:2015uuo,Hall:2015xds,Wang:2015omi,Salvio:2015jgu,Son:2015vfl,Cai:2015hzc,Bizot:2015qqo,Hamada:2015skp,Kang:2015roj,Jiang:2015oms,Jung:2015etr,Gu:2015lxj,Goertz:2015nkp,Ko:2016lai,Palti:2016kew,Karozas:2016hcp,Bhattacharya:2016lyg,Cao:2016udb,Faraggi:2016xnm,Han:2016bvl,Kawamura:2016idj,King:2016wep,Nomura:2016rjf,Harigaya:2016pnu,Han:2016fli,Hamada:2016vwk}
  appeared that account for the diphoton excess with vector-like
  exotics that couple to the $750$ GeV particle.}, we will extend the
analysis of the previous section by taking into account additional
exotics allowed by the string spectrum and a variety of fermion masses
$M_i$ consistent with current bounds on vector-like quarks. We will
also study the ultraviolet perturbativity of these models (i.e.,
assuming a large string scale), which can present an issue in some
models with vector-like exotics
\cite{Dhuria:2015ufo,Gu:2015lxj,Zhang:2015uuo,Salvio:2015jgu,Goertz:2015nkp,Hamada:2015skp,Bae:2016xni,Hamada:2016vwk}.

\subsection{Renormalization Group Equations and Infrared Fixed Points}

We first consider whether the model can remain perturbative up to a
large ultraviolet string scale, e.g., $\mathcal{O}(10^{16} \text{
  GeV})$. For definiteness, we assume that the theory is
supersymmetric down to the TeV scale, but similar conclusions would
hold for a larger breaking scale, or even without supersymmetry.

We consider models where the $750$ GeV
particle is a scalar degree of freedom in a chiral supermultiplet $S$
that couples to $N_i$ exotic vector-like chiral multiplets $X_i,\ov X_i$ in the superpotential
via a Yukawa coupling $\gamma_i\, S X_i \ov X_i$. The exotics $X_i$ and $\ov X_i$ transform
as $(n_3^i, n_2^i)_{y_i}$ and  $(n^{i\ast}_3, n^{i\ast}_2)_{-y_i}$, respectively, with $q_i=t_{3,i}+y_i$.

In supersymmetric models the gauge couplings are governed by the
renormalization group equations
\begin{align}
16 \pi^2 \beta_{g_3} & = \left[ -3 + 2 \sum_i N_i T^3(i) n^i_2 \right] g_3^3 \nonumber \\
16 \pi^2 \beta_{g_2} & = \left[ +1 +2 \sum_i N_i T^2(i) n^i_3 \right] g_2^3 \nonumber \\
16 \pi^2 \beta_{g_1} & = \left[ +\frac{33}{5} + 2 \sum_i N_i T^1(i) n^i_2 n^i_3 \right] g_1^3 .
\end{align}
where $\beta_{g_i} \equiv dg_i/dt$ with $t=\ln(\mu/\mu_0)$. $T^a(i)$ is
the Dynkin index for representation $i$ in group $a$. The Dynkin
indices of low-dimensional representations are $T^3(i) = (3,
1/2, 0)$ for $n^i_3=(8, 3, 1)$, $T^2(i) = (2, 1/2, 0)$ for $n^i_2=(3,
2, 1)$, and $T^1(i) \equiv \frac{3}{5} y_i^2$.   We have
used the GUT-normalized gauge coupling $g_1$ for $U(1)_Y$, which is
related to th ordinary $g'$ by $g_1=\sqrt{5/3} g'$. The beta functions for
the Yukawa coupling $\gamma_i$ are [see, e.g., \cite{Martin:1997ns}]
\begin{equation}
16 \pi^2 \beta_{\gamma_i} = 2 \gamma_i |\gamma_i|^2 +  \gamma_i \left( \sum_j N_j n^j_2 n^j_3 |\gamma_j|^2\right) 
 -4 \gamma_i \sum_{a=1}^3 C_2^a(i) g_a^2,
\end{equation}
  where $C_2^i(a)$ is the quadratic Casimir: $C_2^3(i) = (3, 4/3, 0)$ for $n^i_3=(8, 3, 1)$;
 $C_2^2(i) = (2, 3/4, 0)$ for $n^i_2=(3, 2, 1)$; and $C_2^1(i)= T^1(i)= \frac{3}{5} y_i^2$.

The specific set of models we will study have the MSSM spectrum (which can optionally include three right- handed neutrinos)
augmented by $N_Q$ $(3,2)_{1/6}+(\ov 3,2)_{-1/6}$ pairs, $N_U$ $(3,1)_{2/3}+(\ov 3,1)_{-2/3}$ pairs,
$N_D$ $(3,1)_{-1/3}+(\ov 3, 1)_{1/3}$ pairs, $N_L$ $(1,2)_{-1/2}+(1,2)_{1/2}$ pairs, and $N_E$ $(1,1)_1+(1,1)_{-1}$
pairs, all of which occur frequently in the type IIA compactifications \cite{Cvetic:2015vit,Cvetic:2011iq}. The subscripts denote that one of the chiral multiplets in the pair has the same MSSM quantum numbers as
the associated MSSM superfield, e.g., $Q, U, D, L, E$. The Yukawa couplings in this model
will be labeled similarly, i.e., $\gamma_Q, \gamma_U, \gamma_D, \gamma_L, \gamma_E$. The beta functions
for the gauge couplings are
\begin{align}
16 \pi^2 \beta_{g_3} &= g_3^3(-3+2N_Q+N_U+N_D) \nonumber \\
16 \pi^2 \beta_{g_2} &= g_2^3(1+3N_Q+N_L) \nonumber \\
16 \pi^2 \beta_{g_1} &= g_1^3\left(\frac{33}{5} + \frac65 \left(\frac{N_Q}{6}+\frac{4N_U}{3}+\frac{N_D}{3}+\frac{N_L}{2}+N_E\right) \right). 
\end{align}
Those for the Yukawa couplings are
\begin{align}
16 \pi^2 \beta_{\gamma_Q} &= \gamma_Q \left[2|\gamma_Q|^2 + \alpha -4\left(\frac43 g_3^2+\frac34 g_2^2+\frac35\left(\frac16\right)^2g_1^2\right) \right] \nonumber \\
16 \pi^2 \beta_{\gamma_U} &= \gamma_U \left[2|\gamma_U|^2 + \alpha -4\left(\frac43 g_3^2+\frac35 \left(\frac23\right)^2g_1^2\right) \right] \nonumber \\
16 \pi^2 \beta_{\gamma_D} &= \gamma_D \left[2|\gamma_D|^2 + \alpha -4\left(\frac43 g_3^2+\frac35 \left(\frac13\right)^2g_1^2\right) \right] \nonumber \\
16 \pi^2 \beta_{\gamma_L} &= \gamma_L \left[2|\gamma_L|^2 + \alpha -4\left(\frac34 g_2^2+\frac35 \left(\frac12\right)^2g_1^2\right) \right] \nonumber \\
16 \pi^2 \beta_{\gamma_E} &= \gamma_E \left[2|\gamma_E|^2 + \alpha -4\left(\frac35 g_1^2\right) \right], \label{betas}
\end{align}
where 
\begin{equation}
\alpha = 6N_Q|\gamma_Q|^2+3N_U|\gamma_U|^2+3N_D|\gamma_D|^2+2N_L|\gamma_L|^2+N_E|\gamma_E|^2.
\end{equation}
We will study models with specific values for the tuple $(N_Q,N_U,N_D,N_L,N_E)$.

\vspace{1cm}

The perturbative nature of specific models is ensured in part by the
existence of infrared fixed points. For reasonable
ultraviolet boundary conditions for the Yukawa couplings in the range
$[0.1,10]$, Yukawa couplings of vector-like quarks
often approach their fixed points. 

Let us first justify this range of 
UV Yukawa couplings in type IIA compactifications with intersecting D6-branes, which are one of the contexts
for our previous work \cite{Cvetic:2015vit}. We shall focus on the allowed
magnitudes of the Yukawa couplings at the string scale.  These were
calculated exactly at the string (world-sheet) tree level for toroidal
compactifications in \cite{Cvetic:2003ch} (for related works see,  \cite{Cremades:2003qj,Lust:2004cx,Cremades:2004wa}).  The full
expression (both classical and quantum part of the string tree level
amplitude) for branes wrapping factorizable three-cycles of $T^6=T^2\times
T^2\times T^2$ is written as (see, eq. (3) of \cite{Cvetic:2003ch}):
\begin{equation}
\gamma=\sqrt{2}g_s 2\pi\prod_{j=1}^3\left[\frac{16\pi^2 B(\nu_j,1-\nu_j)}{
    B(\nu_j,\lambda_j)B(\nu_j,1-\nu_j-\lambda_j)}\right]^\frac14\sum_m\exp\left(-\frac{A_j(m)}{2\pi\alpha'}\,\right)  .
    \end{equation}
Here the chiral superfields are localized at the intersections of pairs of D6-branes, which 
intersect at respective angles $\pi\nu_j$, $\pi \lambda_j$,  and $\pi- \pi\nu_j-\pi\lambda_j$ on the   $j$th two-torus.
$A_j(m)$ is the area of the triangle formed by the three
    intersecting D6-branes on the j-th two-torus and  $g_s=e^{\Phi/2}$, with $\Phi$ corresponding to the Type IIA dilaton.  The beta function $B(p,q)$ is defined in terms of $\Gamma(p)$ functions as $B(p,q)\equiv \frac{\Gamma(p)\, \Gamma(q)}{\Gamma(p+q)}$.
The coupling is between two fermion  fields and a scalar field, i.e., the
massless states appearing at the respective intersections,  whose
kinetic energies are taken to be canonically normalized.

\begin{figure}[tb]
\begin{center}
\includegraphics[scale=.6]{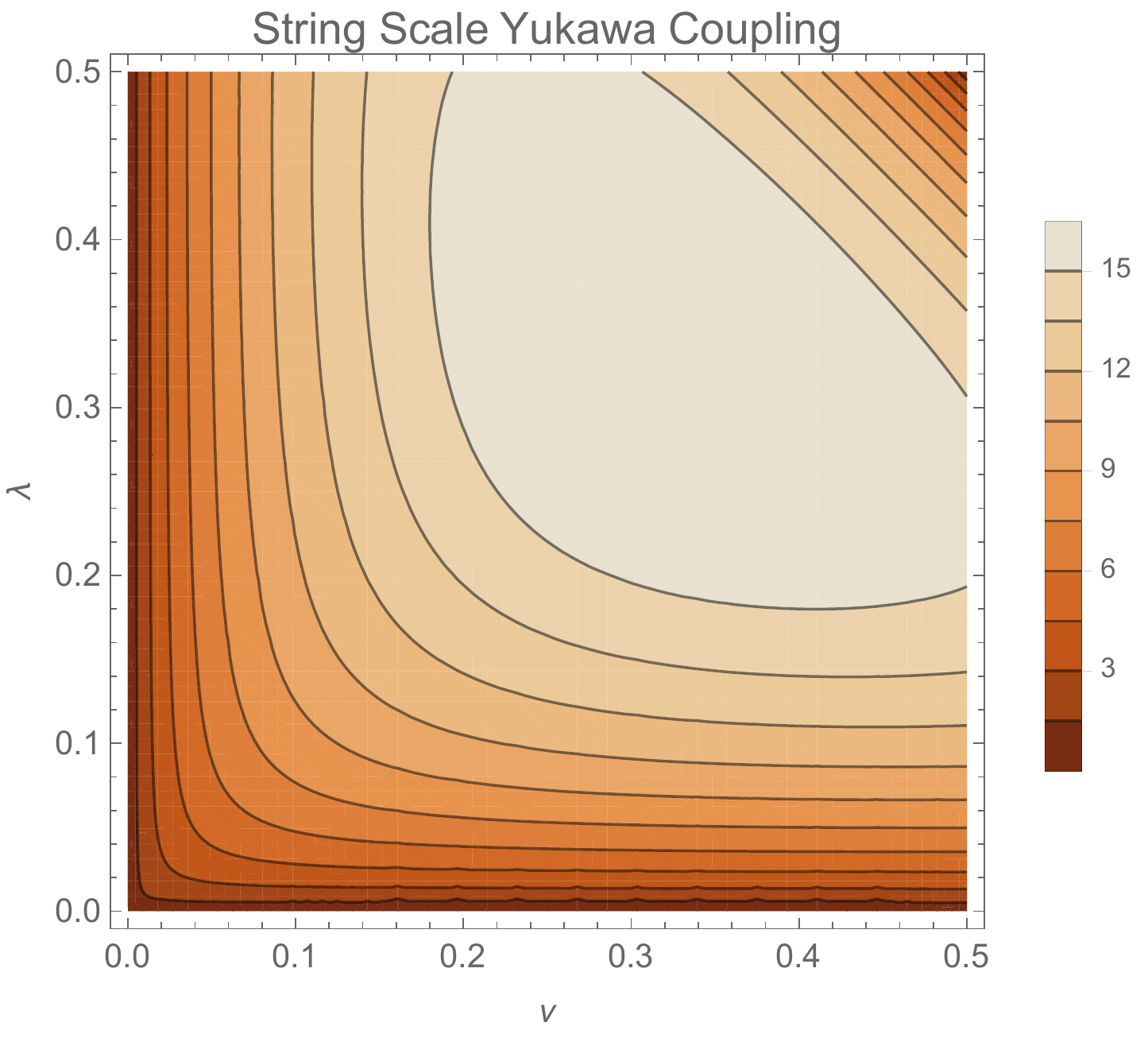}
\end{center}
\caption{String scale Yukawa coupling for $g_s=0.2$ and brane angles $\lambda, \nu$.}
\label{fig:UVYukawa}
\end{figure}

The magnitude of Yukawa couplings can be surprisingly
sizable, reaching ${\cal O}(10)$, while having the string coupling
still perturbative. The choice of brane angles and disc instanton
areas that maximize the Yukawa coupling are
$\nu_j=\lambda_j=1-\nu_j-\lambda_j=\frac{1}{3}$ ($j=1,2,3$) and
$A_{j}(m=1) =0$, respectively. Taking these values and a perturbative
string coupling $g_s=0.2$, the Yukawa coupling is $\gamma=17$. Since
$\gamma\propto g_s$ the Yukawa coupling can be even higher while
remaining within a perturbative string framework. Taking universal
angles for all tori $\nu_j=:\nu$ and $\lambda_j=:\lambda$, the
dependence of $\gamma$ on these angles is presented in Figure
\ref{fig:UVYukawa}, which demonstrates that for zero disc instanton
area the Yukawa couplings $\gamma$ are $>1$ for a wide variety of angles.
In summary, perturbative Type IIA string theory therefore allows for a
range of $\gamma$ including large values in the interval $\gamma\in [{\cal O}(1),\ {\cal
  O}(50)]$.

We now turn to an approximate analytic analysis of the range of Yukawa
couplings in the IR regime, in particular for $\gamma_Q$.  We will
demonstrate that for a broad range of UV boundary conditions
$\gamma_Q$ robustly approaches an approximate IR fixed point, governed
by the IR value of $g_3$, the largest gauge coupling.  (Note, for
example, an early analysis of such an IR behavior for the fourth
family Yukawa couplings within the MSSM \cite{Cvetic:1985fp}.)

First one observes from (\ref{betas})  that $\gamma_L$  tends to decrease in the IR regime due to a positive, dominant contribution from $\gamma_Q$ and a smaller, negative contributions from $g_2$. We shall reconfirm post-factum that  for $\gamma_L ={\cal O}(1)$ in the UV,  $\gamma_L < \gamma_Q$ in the IR.  

To illustrate the IR fixed point behavior, let us study the beta
function for $\gamma_Q$ in (\ref{betas}) in the case of only $Q$
exotics, i.e., $N_Q\ne 0$ and $N_{U,D,L,E}=0$. For simplicity we
neglect $g_{1,2}$ relative to $g_3$ and replace the running $g_{3}$
with its (approximately ``constant'') IR value at $\Lambda_{IR}\sim 1$
TeV. This approximation is justified since the gauge couplings run
logarithmically with the scale $\Lambda$, while the IR fixed point for
Yukawa couplings is approached with a power-law for $\Lambda$.  With
these approximations we obtain:
\begin{equation}
16 \pi^2 \beta_{\gamma_Q} = \gamma_Q \left[2(1+3N_Q)\gamma_Q^2 - \frac{16}{3} g_{3\, IR}^2 \right] \, ,
\end{equation} 
which is easily solved to yield
\begin{equation}
\gamma_{Q\, IR}^2= \frac{a}
{\left[1-\left(1-\frac{a}{\gamma_{Q\, UV}^2}\right) \left(\frac{\Lambda_{IR}}{\Lambda_{UV}}\right)^{\frac{2a}{b}}\right]},
\label{eqn:solvedyukawa}
\end{equation}
where $a= \frac{16}{3} \frac{g_{3\, IR}^2}{2(1+3N_Q)}$ and $b=\frac{16\pi^2}{2(1+3N_Q)}$.
We thus observe an IR  robust fixed point governed  by $a$. Although \eqref{eqn:solvedyukawa} gives a reasonable
approximation to $\gamma_Q$, we will use the exact solutions to \eqref{betas} in our subsequent
analysis.

\subsection{Perturbative and Stable Models of the Diphoton Excess}

\begin{figure}[tb]
\begin{center}
\includegraphics[scale=.3]{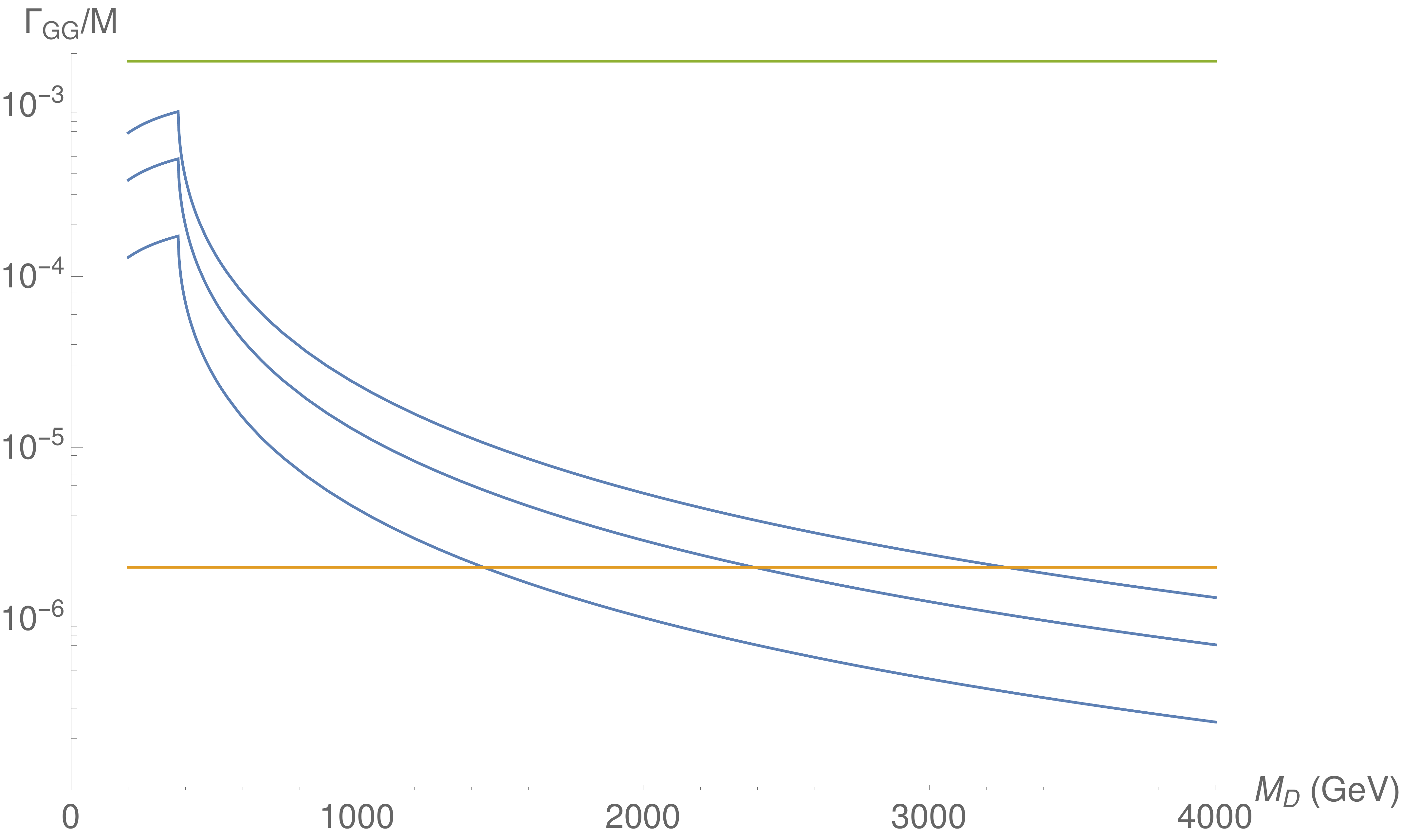}
\end{center}
\caption{$\Gamma_{GG}/M$ as a function of $M_D$ for $N_D=N_L=3,2,1$ are the top, middle, and bottom blue lines, respectively. Upper
and lower bounds on the rate from \cite{Franceschini:2015kwy} are also given.}
\label{fig:GamGG}
\end{figure}

The existing experimental bounds on new vector-like fermions are very
model dependent. Assuming decays into standard model particles such as
$D\rightarrow W t, Z b,$ or $ Hb$ the current 95\% C.L. lower limits
are in the range 740-900 GeV~\cite{Khachatryan:2015gza} or 575-813
GeV~\cite{Aad:2015kqa} for CMS and ATLAS, respectively.  The
corresponding limits for a heavy charge-2/3 quark are 720-920
GeV~\cite{Khachatryan:2015oba} and 715-950 GeV~\cite{Aad:2015kqa}.
Those for charged and neutral leptons are much weaker, typically
around 100 GeV~\cite{Agashe:2014kda}, although some mass ranges up to
$\sim$180 GeV are excluded \cite{Aad:2015dha}. We will simply assume
$M_i \gtrsim 750$ GeV (quarks) and $\gtrsim 200$ GeV (leptons).

\begin{figure}[htb]
\begin{center}
\makebox[\textwidth][c]{
\includegraphics[scale=.4]{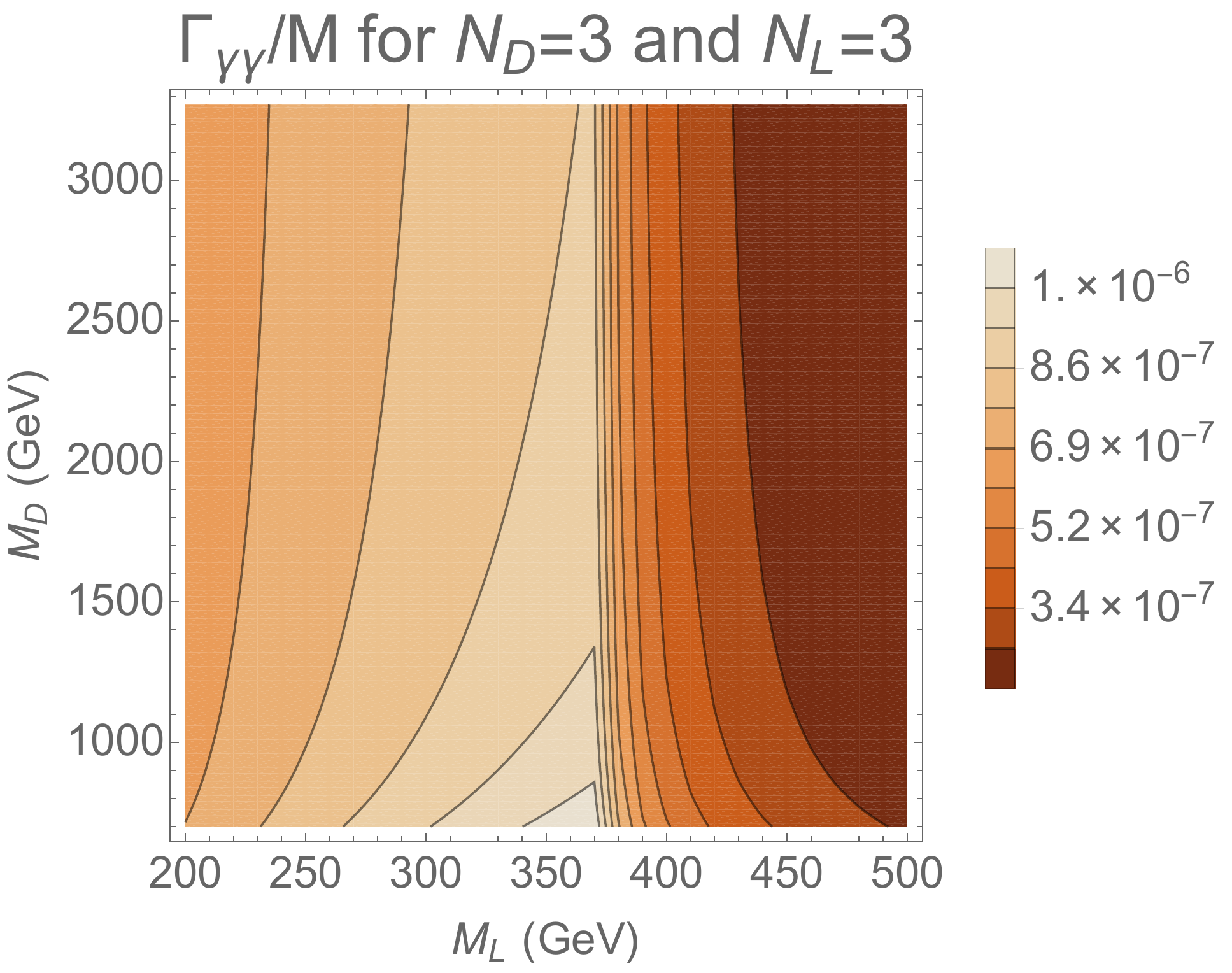}
\includegraphics[scale=.4]{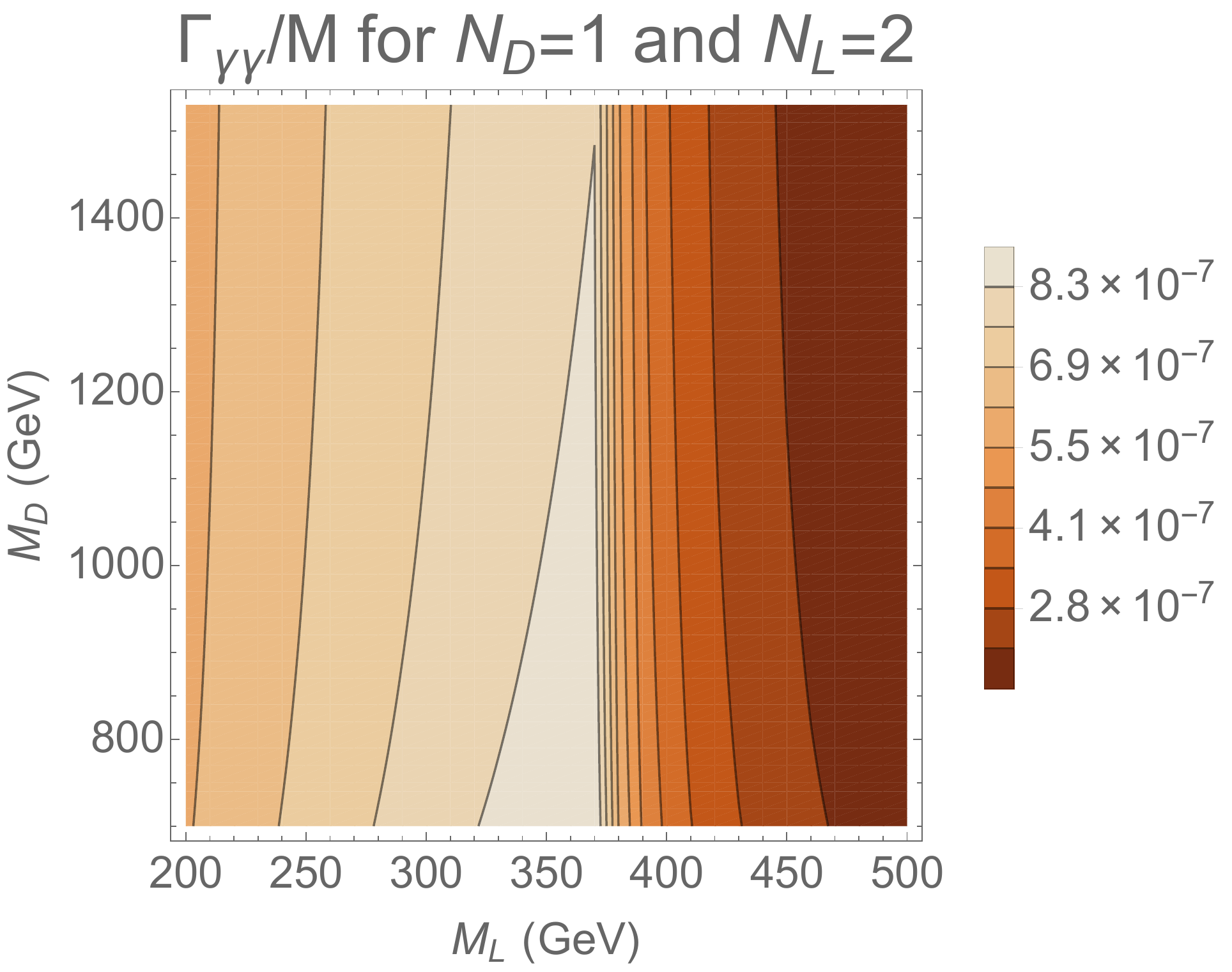}
}
\end{center}
\caption{The partial width into photons for models with $(N_D,N_L)=(3,3)$ and $(1,2)$.}
\label{fig:33and12}
\end{figure}

We consider models with $N_L$ vector-like lepton doublets and $N_D$
down-type quark chiral supermultiplets that couple in the
superpotential to an MSSM singlet $S$ that contains a pseudoscalar $s$
of mass $M_s=750$ GeV. We choose $t_{UV}=30$ $(t_{IR}=0)$ relative to
a reference scale $\mu_0=750$ GeV, corresponding to
$\mu_{UV}=8.0\times 10^{15}$ GeV ($\mu_{IR}=750$ GeV).  We assume
universal ultraviolet Yukawa couplings with $\gamma_{UV}=1$, but have
verified that larger values give almost identical results because of
the IR fixed point. We take initial values $(\alpha_3, \alpha_2,
\alpha_1)=(0.092, 0.033, 0.017)$ at 750 GeV, obtained by running from
$M_Z$ using the SM $\beta$ functions. Threshold corrections associated
with $M_D$ and $M_L$ or for a larger supersymmetry breaking scale,
e.g., up to tens of TeV, are small. We assume universal masses $M_L$
and $M_D$ for the vector-like leptons and quarks.  

The partial width $\Gamma(s\to GG)$ must satisfy $\Gamma_{GG}/M\geq
8\times 10^{-7}$ to be in the preferred blue band on the left side of
Figure $1$ in \cite{Franceschini:2015kwy}.  We study
$\Gamma_{GG}/M\geq 2\times 10^{-6}$, for which the blue band flattens
out and the analysis of the rate into two photons is simplified. For
$N_D=N_L=3,2,1$ the rate $\Gamma_{GG}$ is computed in Figure
\ref{fig:GamGG} as a function of the exotic quark mass
$M_D$. $\Gamma_{GG}/M\geq 2\times 10^{-6}$ for $M_D\lesssim 3270,
2380, 1430$ GeV, respectively, well within vector-like quark
bounds. These exotic representations embed into $5 + 5^\ast$ pairs,
which maintain MSSM-like gauge unification to lowest order, and
$N_D=N_L=3$ is motivated by $E_6$ models.

For $\Gamma_{GG}/M \geq 2\times 10^{-6}$, the partial width into
photons must satisfy \cite{Franceschini:2015kwy} $6\times
10^{-7}\lesssim \Gamma_{\gamma\gamma}/M \lesssim 2\times10^{-6}$ at
2$\sigma$, assuming no other contributions to the width. The $E_6$
motivated model $(N_D,N_L)=(3,3)$ and the minimal model that can
account for the data $(N_D,N_L)=(1,2)$ are presented in Figure
\ref{fig:33and12}.  In each plot $M_D$ goes up to the maximal value
that allows for $\Gamma_{GG}/M\geq 2\times10^{-6}$. In both cases
obtaining a large enough $\Gamma_{\gamma\gamma}/M$ requires
$M_L\lesssim 375$ GeV, while obtaining a large enough $\Gamma_{GG}/M$
requires $M_D\lesssim 3.3, 1.6$ TeV, respectively.

Similar $\Gamma_{\gamma\gamma}/M$ plots for all $N_D,N_L\in \{1,2,3\}$ are presented in Figures \ref{fig:nd3}, \ref{fig:nd2}, and \ref{fig:nd1}.
Interestingly, $\Gamma_{\gamma\gamma}/M$ tends to increase with decreasing $N_D$ for fixed $N_L$.

\section{Conclusions}
One  possible interpretation of the 750 GeV diphoton excess is a new scalar or pseudoscalar resonance, coupled to gluons and photons through loops of exotic vector-like fermions. Such scalars and  exotics are common in string theory, and we have previously argued that they are required by tadpole cancellation conditions in many weakly-coupled intersecting brane constructions. In this paper we reexamined the dataset of D-brane quivers that contain the MSSM and found that many of them indeed contain standard model singlets that could be candidates for the 750 GeV particle, as well as having perturbatively-allowed Yukawa couplings to exotic vector-like fermions. Only certain quantum numbers for those exotics are common, especially left and right-chiral pairs of $SU(2)$-doublet leptons with electric charges $0$ and $-1$, $SU(2)$-singlet leptons with  charge  $-1$,  and  $SU(2)$-singlet down-type quarks with charge $-1/3$. Up-type $SU(2)$-singlets with charge $2/3$
and doublets with charges $2/3$ and $-1/3$ also occur, but less frequently. In each case the pairs are non-chiral with respect to an additional perturbative global symmetry that prevents them from obtaining a string-scale mass.

Following the phenomenological analysis in \cite{Franceschini:2015kwy}, we showed that the diphoton excess could be accounted for in this theoretical framework with a large ultraviolet string scale (e.g., $10^{16}$ GeV) provided that it is narrow, i.e., the width is mainly due to $GG$ and $\gamma\gamma$. We argued that even though the allowed Yukawa couplings could be very large (of $\mathcal{O}(10)$ at the string scale, they are typically driven to lower values $\lesssim 1$ at low energies due to an IR fixed point. For a low enough number of exotics, the gauge and Yukawa couplings can remain perturbative up to the UV scale (and even be consistent with MSSM-type gauge unification at one loop). The observed diphoton rate
can be reproduced for pseudoscalar couplings to fermions for relatively light vector-like lepton pairs, with masses close to 375 GeV, and heavier vector-like quark pairs with masses up to around 3.3 TeV, all safely within present experimental
limits but still in the range that could be observed at the LHC. Scalar couplings to fermions give  lower rates unless they are enhanced by spin-0 particles in the loops, which involve unknown $A$-term coefficients in the supersymmetric case.
These conclusions are insensitive to the supersymmetry-breaking scale, or even whether supersymmetry survives to low energies.

On the other hand, it would be difficult to accommodate the large ($\Gamma/M\sim 0.06$) width suggested by the ATLAS data in this framework. One would require either  large Yukawa couplings in the IR or a large number of exotics, either of which would lead to strong coupling at the TeV scale. (Other possibilities suggested by many other authors include different event topologies, or the existence of two resonances separated by 10's of GeV, such as the scalar and pseudoscalar components of a complex scalar.)

\acknowledgments J.H. thanks Brent Nelson for discussions.  
This research is supported in part by the DOE Grant
Award DE-SC0013528, (M.C.), the Fay R. and Eugene L. Langberg Endowed
Chair (M.C.) and the Slovenian Research Agency (ARRS) (M.C.). 
J.H. is supported by startup funding from Northeastern University.

\appendix

\begin{figure}[htb]
\begin{center}
\makebox[\textwidth][c]{
\includegraphics[scale=.4]{dl33.pdf}
\includegraphics[scale=.4]{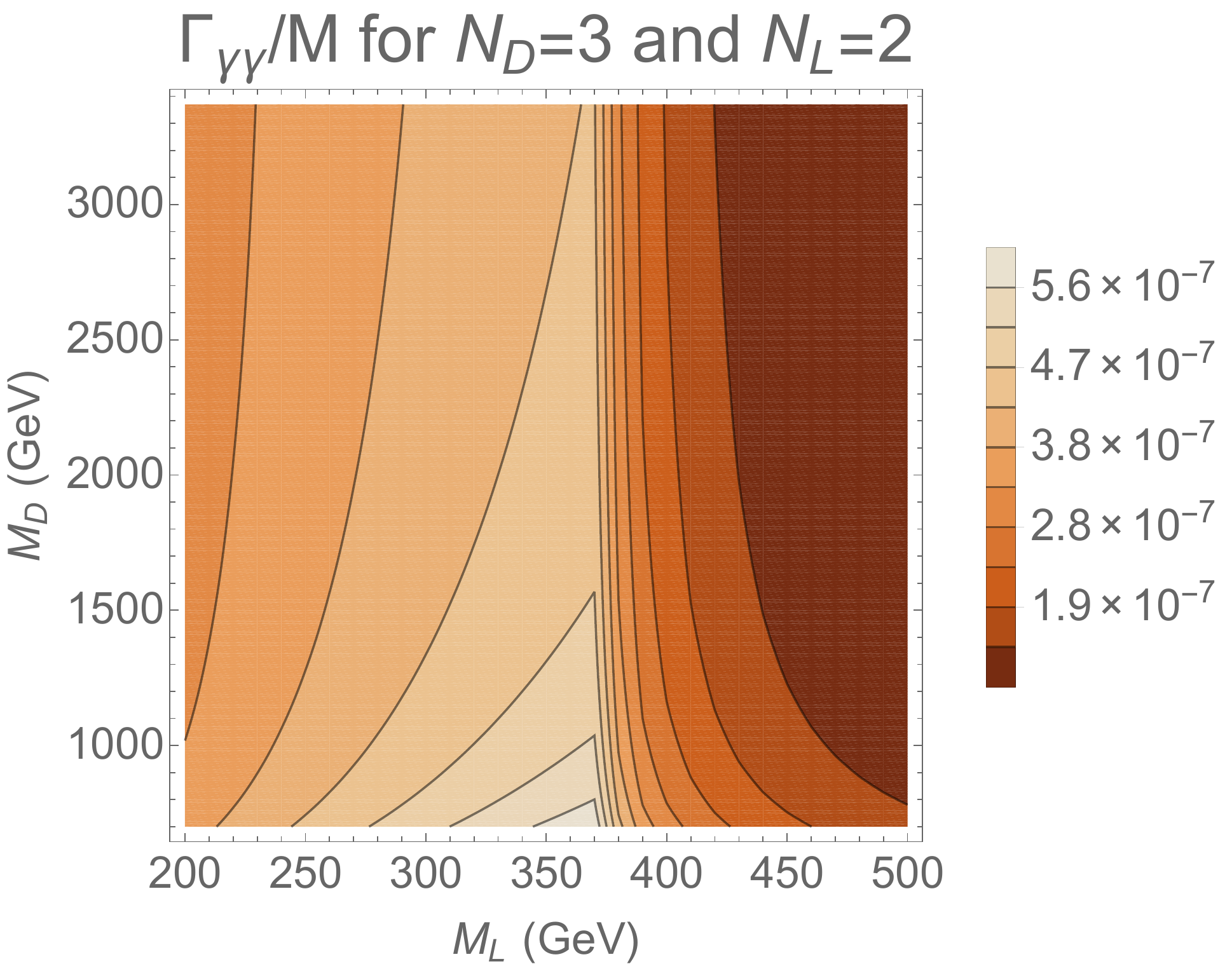}
}
\makebox[\textwidth][c]{
\includegraphics[scale=.4]{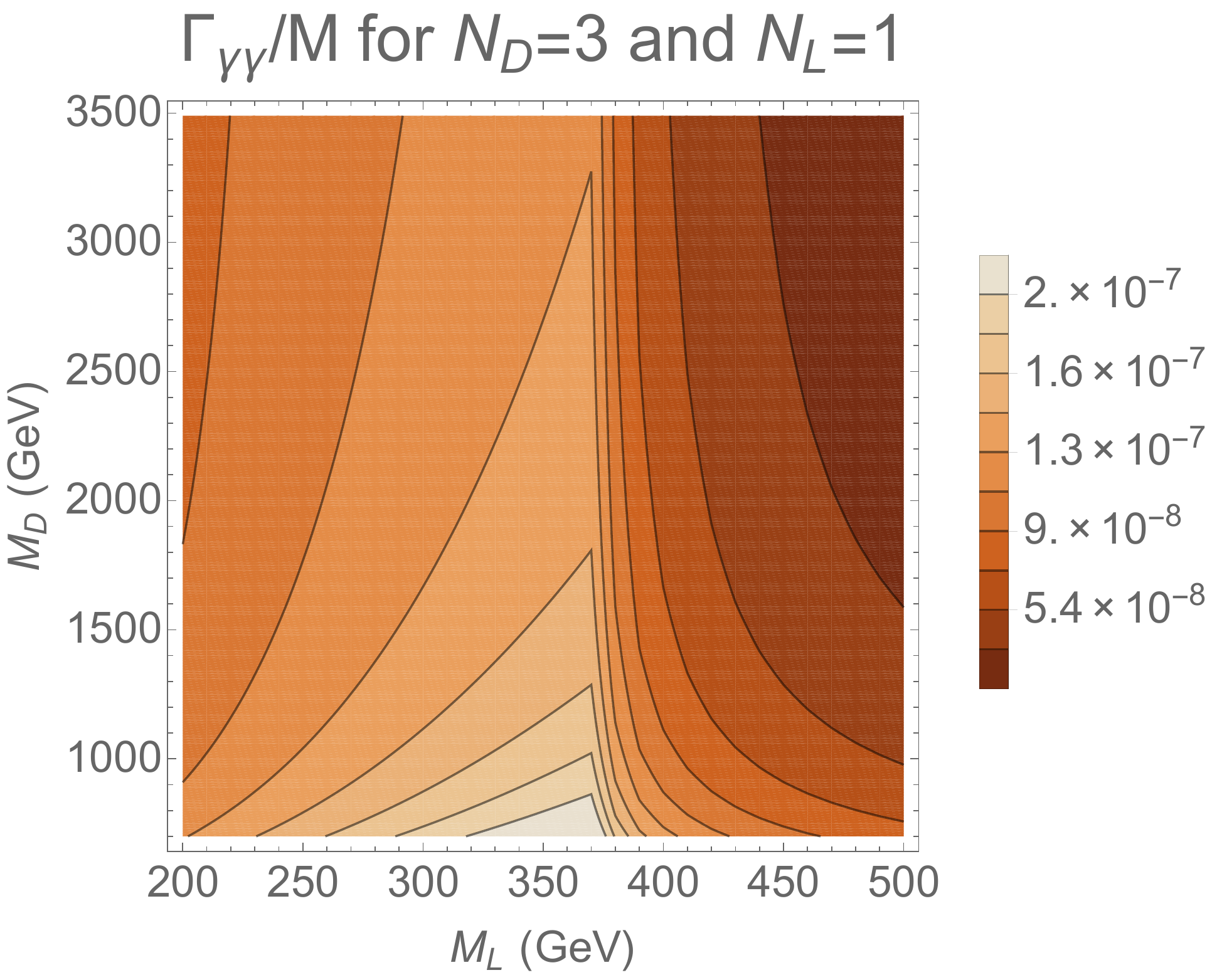}
}
\end{center}
\caption{The partial width into photons for models with $N_D=3$ and $N_L=3,2,1$.}
\label{fig:nd3}
\end{figure}

\begin{figure}[htb]
\begin{center}
\makebox[\textwidth][c]{
\includegraphics[scale=.4]{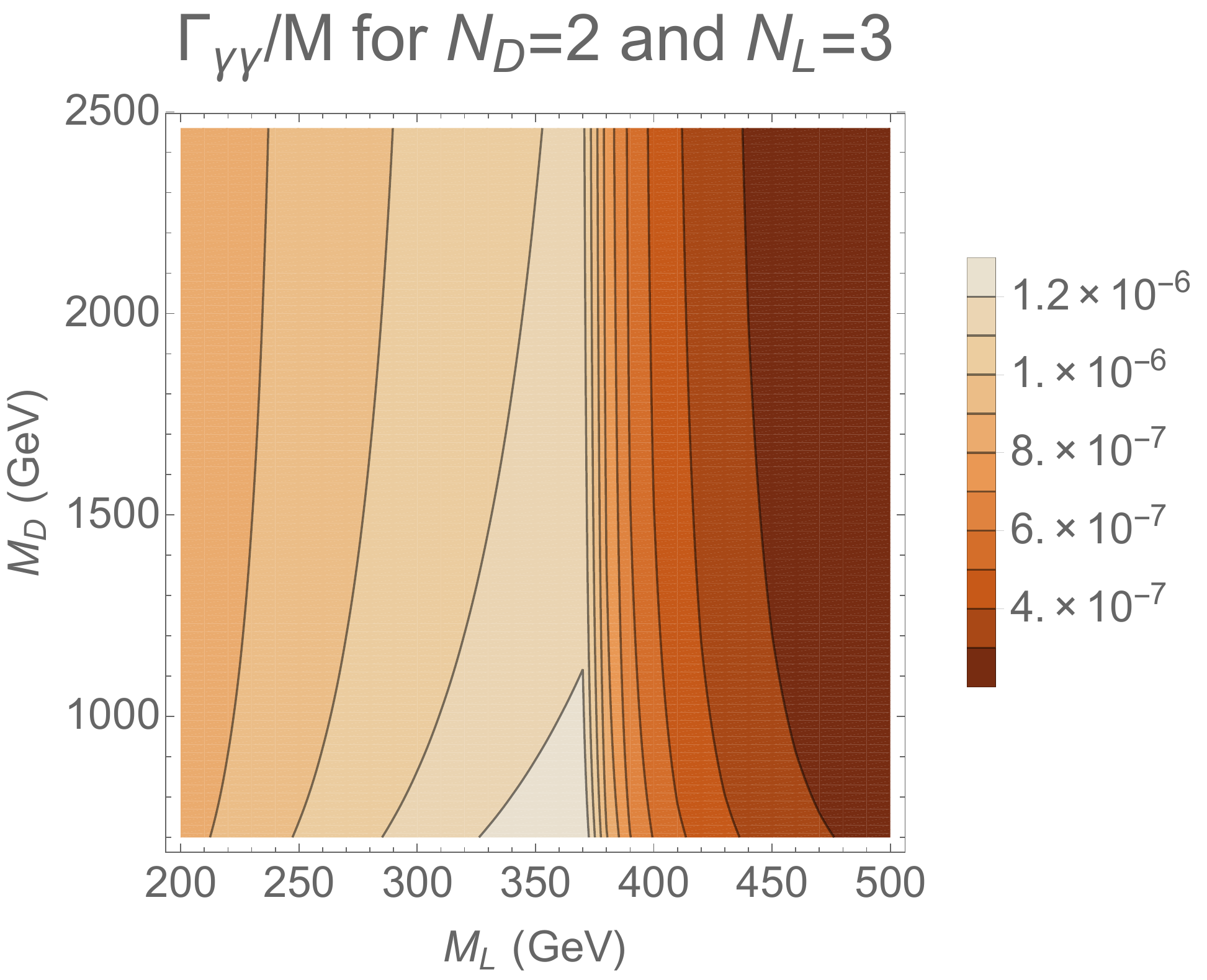}
\includegraphics[scale=.4]{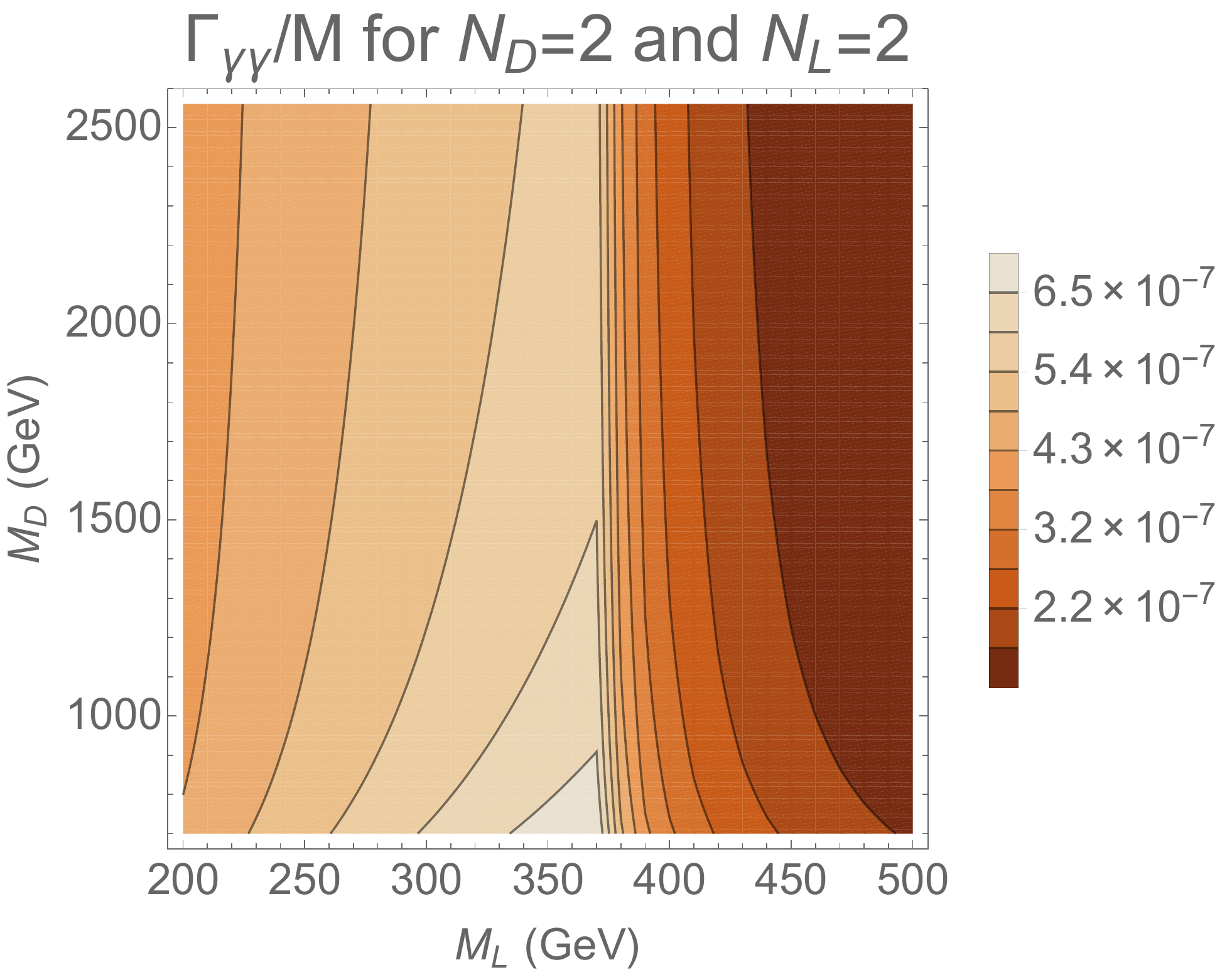}
}
\makebox[\textwidth][c]{
\includegraphics[scale=.4]{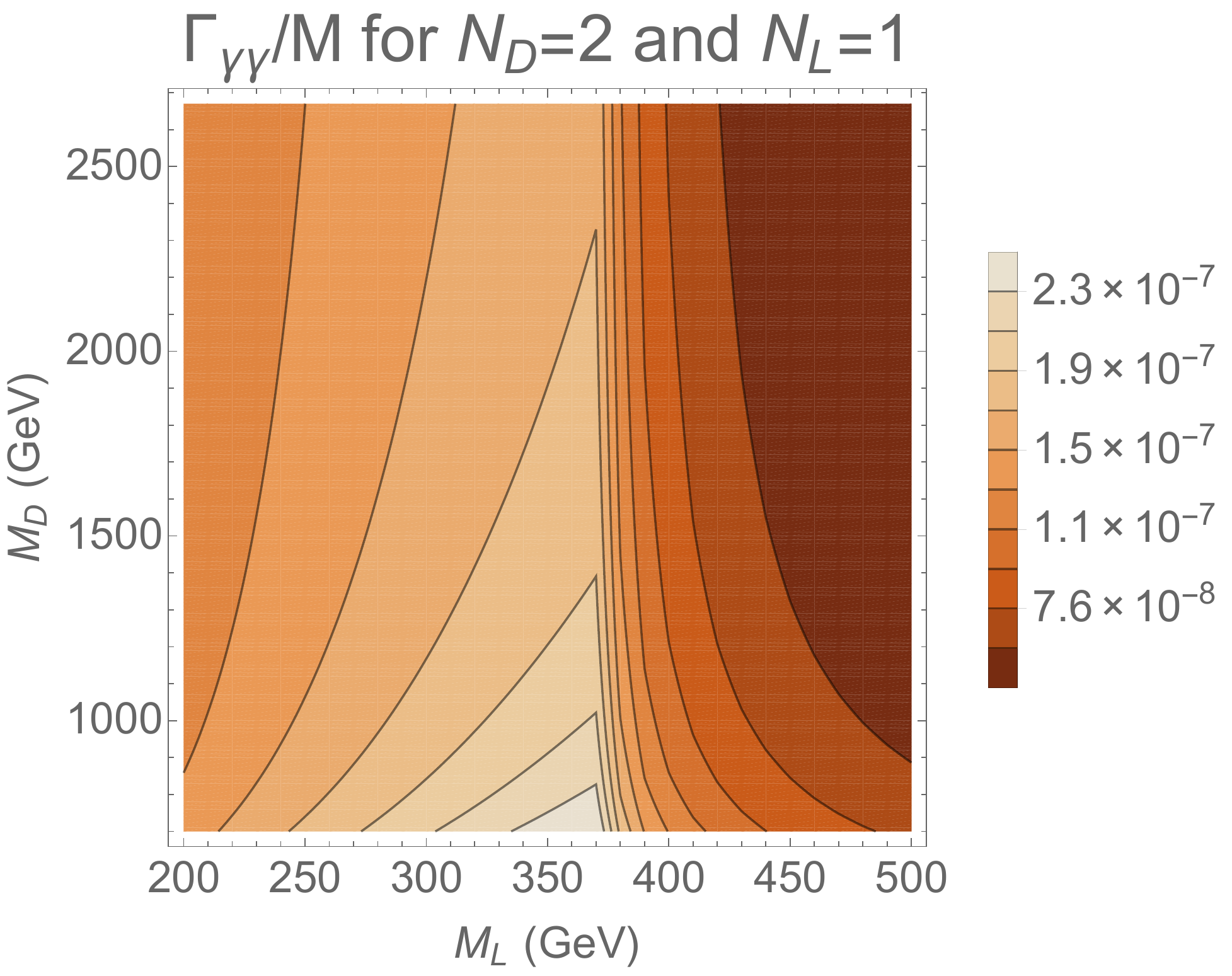}
}
\end{center}
\caption{The partial width into photons for models with $N_D=2$ and $N_L=3,2,1$.}
\label{fig:nd2}
\end{figure}

\begin{figure}[htb]
\begin{center}
\makebox[\textwidth][c]{
\includegraphics[scale=.4]{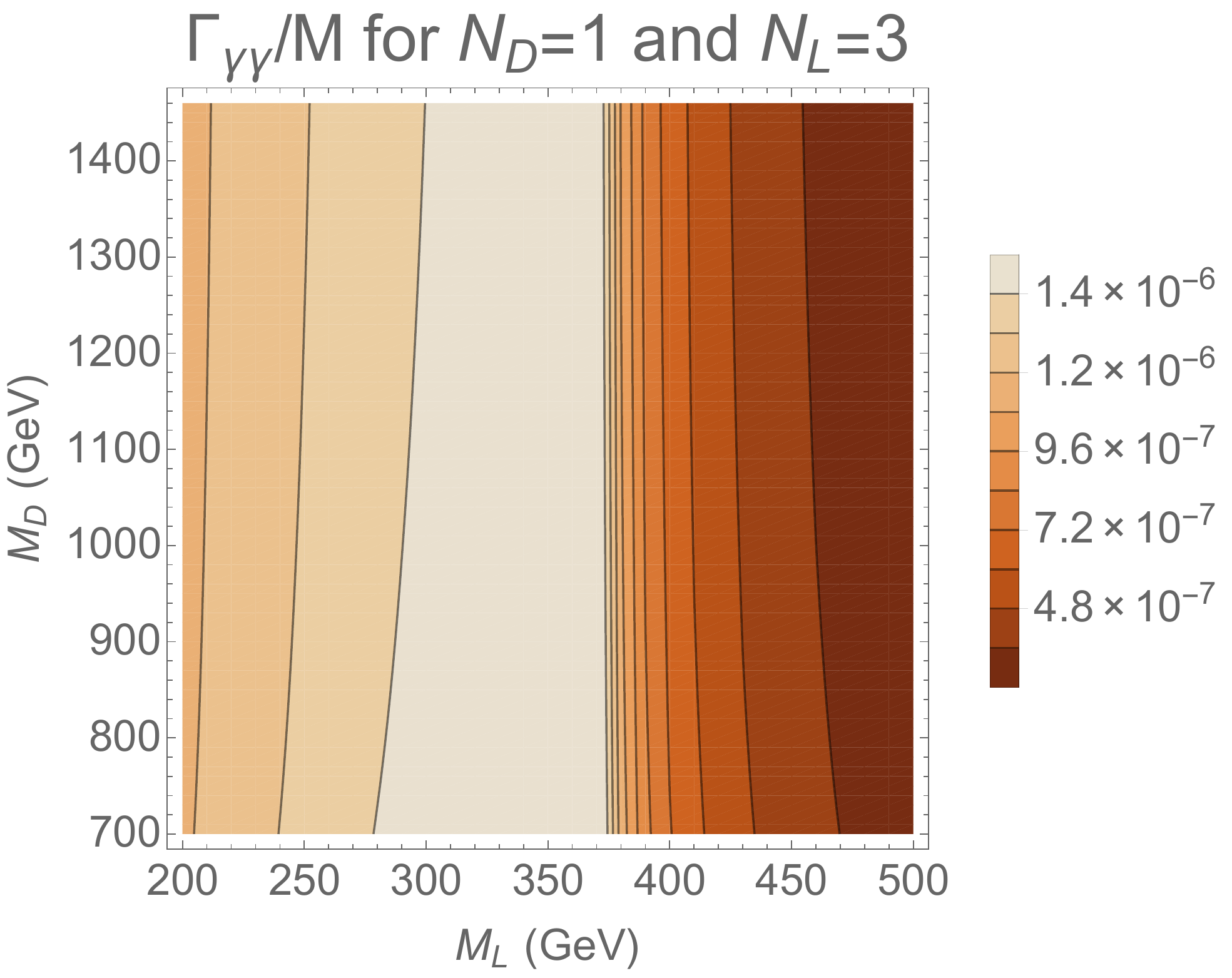}
\includegraphics[scale=.4]{dl12.pdf}
}
\makebox[\textwidth][c]{
\includegraphics[scale=.4]{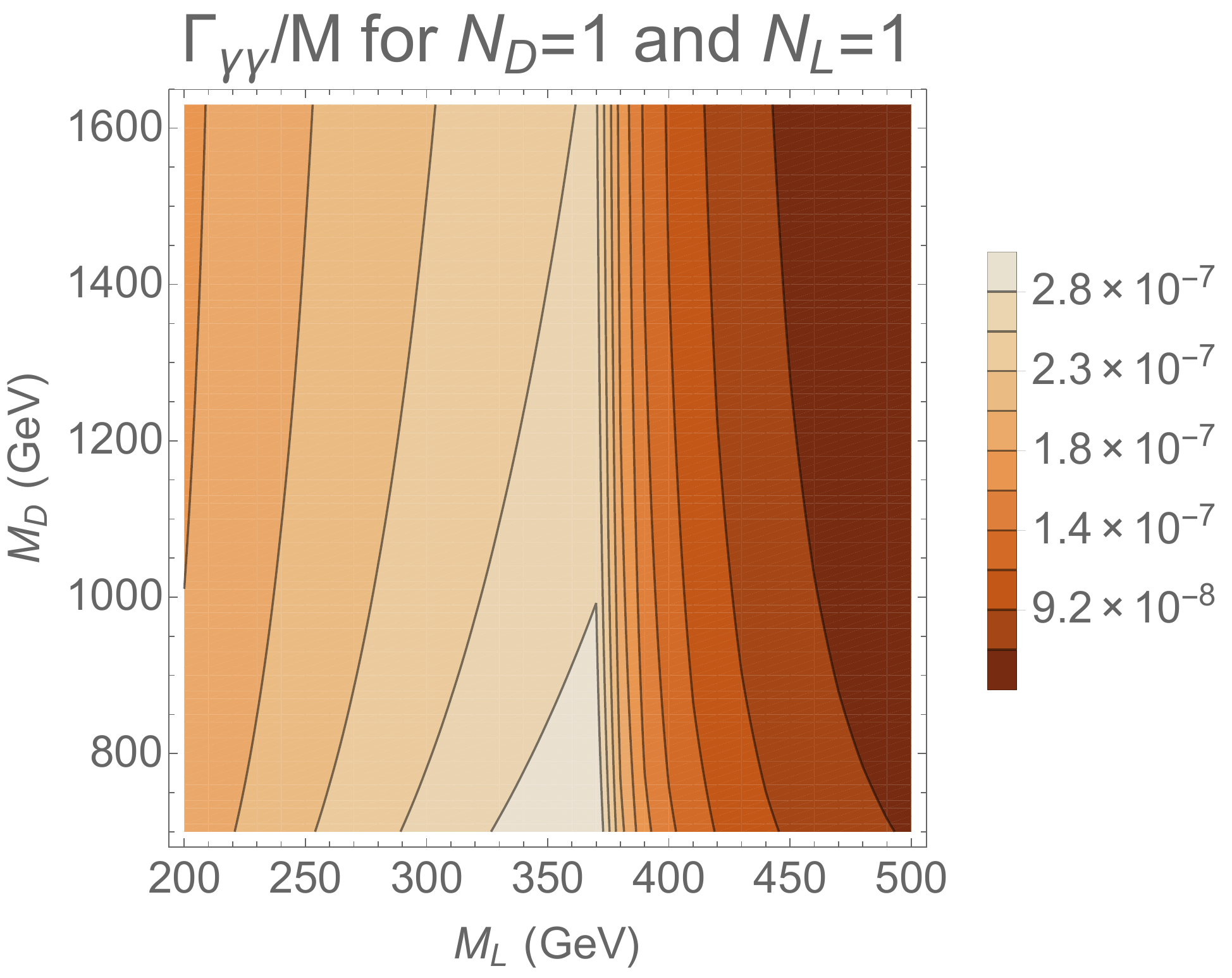}
}
\end{center}
\caption{The partial width into photons for models with $N_D=1$ and $N_L=3,2,1$.}
\label{fig:nd1}
\end{figure}

\section{Weakly coupled type II compactifications with orientifolds and D-branes}
\label{sec:D-brane appendix}

We will now review basic facts about weakly coupled
compactifications with intersecting D-branes and orientifolds that
provided the context for the rest of the paper.

Compactifications of the type IIa or IIb superstring on a Calabi-Yau
threefold $X$ preserve $\cN=1$ supersymmetry in four dimensions. We 
consider such a context even though our analysis focuses more on
possible exotic sectors than the presence or absence of supersymmetry
at the weak scale. For the sake of concreteness we use the
language of intersecting D6-brane models in the type IIa theory, even
though the results apply equally well in the type IIb theory with
intersecting D7-branes or the type I theory with D9-branes. One
advantage of the type IIa theory is that the chiral spectrum of the
theory is geometrically determined by the intersections of the
D6-branes rather than an interplay between brane intersections and
worldvolume fluxes as in the type IIb and type I cases.

In the type IIa theory a stack of $N$ D6-branes on a special Lagrangian
three-cycle $\pi$ with $[\pi] \in H_3(X,\bZ)$ gives rise to a seven-dimensional
gauge sector on $\bR^{3,1}\times \pi$ that Kaluza-Klein reduces to an $\cN=1$
gauge sector on $\bR^{3,1}$. The theory is equipped with an antiholomorphic
orientifold involution $\sigma$ the fixed-point locus of which is a three-cycle
$\pi_{O6}$ that is wrapped by an O6-plane; under this involution there is also an
orientifold image stack of branes on $\pi'$. If $[\pi]$ is a general class, that is it
does not satisfy special relations with respect to the orientifold, then the $N$
D6-branes on $\pi$ give rise to a $U(N)$ gauge theory; if not, the gauge group may be
symplectic or special orthogonal. If there are a number of D6-brane stacks on general cycles, each with
$N_i$ D6-branes in the stack, then these give rise to a
\begin{equation}
G = \prod_i U(N_i)
\end{equation}
gauge theory, and chiral matter may arise at brane intersections. The
spectrum is determined by the quantization of open strings, and the
spectrum is given in Table \ref{table:intersections} in terms of
topological intersections of D6-brane stacks.

\begin{table}
\centering \vspace{3mm}
\label{table:spectrum}
\begin{tabular}{|c|c|}
\hline
Representation  & Multiplicity \\
\hline $\Yasymm_a$
 & ${1\over 2}\left(\pi_a\circ \pi'_a+ \pi_a\circ \pi_{{\rm O}6}
\right)$  \\
 $\Ysymm_a$
      & ${1\over 2}\left(\pi_a\circ \pi'_a-\pi_a\circ \pi_{{\rm O}6}
\right)$   \\
      $(\fund_a,\antifund_b)$
       & $\pi_a\circ \pi_{b}$   \\
        $(\fund_a,\fund_b)$
	 & $\pi_a\circ \pi'_{b}$
	 \\
	 \hline
	 \end{tabular}
	 \caption{Representations and multiplicities for chiral matter at the intersection of two D6-branes.}
	 \label{table:intersections}
\end{table}

In concrete examples the spectrum is often such that some number of
the $U(1)_i$ diagonal subgroups of $U(N_i)$ is anomalous, in which
case the Chern-Simons couplings of the D-brane give rise to axionic
couplings that cancel the anomaly via the generalized Green-Schwarz
mechanism. These include couplings of the form $\phi F \wedge F$ and
$B \wedge F$ where $\phi$ is the axion, $B$ is its four-dimensional
Hodge dual, and $F$ is a $U(1)$ field strength. The $B\wedge F$
induces a St\"uckelberg mass for the $U(1)$; it is always present if
the $U(1)$ is anomalous, but may also be present when the $U(1)$ is
non-anomalous if associated $\phi F\wedge F$ terms are absent. There
is a condition on the homology of the cycles wrapped by the D6-branes
and O6-plane that is necessary to ensure the absence of certain
$B\wedge F$ type couplings; these are discussed in the introduction.

Finally, D-branes carry Ramond-Ramond charge and source associated flux lines in the extra dimensions. For D-branes
that fill the non-compact spacetime there is an associated Gauss' law, the so-called Ramond-Ramond tadpole
cancellation conditions. In type IIa this is a condition on the homology of cycles wrapped by D6-branes and the O6-plane,
 given by
\begin{equation}
\label{eqn:tadpole}\sum_b N_b \, (\pi_b + \pi_{b'}) = 4 \, \pi_{O6},
\end{equation}
where the sum runs over each of the D-brane stacks. This condition places necessary conditions on the
chiral spectrum that will necessitate the addition of exotics in many examples.

\clearpage
\section{Tables of Dimension 4 Singlet Couplings}
\label{appendix:tables}
\begin{table}[htb]
\begin{center}
\scalebox{.65}{
\centering
\begin{tabular}{|c|c|c|c|} \hline
Exotic Set & Multiplicity & Perturbative $S$ Couplings & Multiplicity  \\ \hline
$(\textbf{1},\textbf{1})_{0}$ $(\textbf{1},\textbf{2})_{-\frac{1}{2}}$ $(\textbf{1},\textbf{2})_{-\frac{1}{2}}$ $(\textbf{1},\textbf{2})_{\frac{1}{2}}$ $(\textbf{1},\textbf{2})_{\frac{1}{2}}$  & $3340$ & & \\
&&  None  & $2404$ \\
&&  $6 \times $ ($(\textbf{1},\textbf{1})_{0}$$(\textbf{1},\textbf{2})_{-\frac{1}{2}}$$(\textbf{1},\textbf{2})_{\frac{1}{2}}$),  & $228$ \\
&&  $4 \times $ ($(\textbf{1},\textbf{1})_{0}$$(\textbf{1},\textbf{2})_{\frac{1}{2}}$$(\textbf{1},\textbf{2})_{-\frac{1}{2}}$),  & $180$ \\
&&  $12 \times $ ($(\textbf{1},\textbf{1})_{0}$$(\textbf{1},\textbf{2})_{\frac{1}{2}}$$(\textbf{1},\textbf{2})_{-\frac{1}{2}}$),  & $128$ \\
&&  $9 \times $ ($(\textbf{1},\textbf{1})_{0}$$(\textbf{1},\textbf{2})_{\frac{1}{2}}$$(\textbf{1},\textbf{2})_{-\frac{1}{2}}$),  & $120$ \\
&&  $1 \times $ ($(\textbf{1},\textbf{1})_{0}$$(\textbf{1},\textbf{2})_{-\frac{1}{2}}$$(\textbf{1},\textbf{2})_{\frac{1}{2}}$),  & $120$ \\
\hline \end{tabular}}
\end{center}
\caption{The multiplicities
of exotic sectors that contain exactly one singlet, binned according to their perturbative singlet couplings in column 3.}
\label{table:One S couplings 1}
\end{table}
\begin{table}[htb]
\begin{center}
\scalebox{.65}{
\centering
\begin{tabular}{|c|c|c|c|} \hline
Exotic Set & Multiplicity & Perturbative $S$ Couplings & Multiplicity  \\ \hline$(\textbf{1},\textbf{3})_{0}$ $(\textbf{1},\textbf{1})_{0}$ $(\textbf{1},\textbf{2})_{-\frac{1}{2}}$ $(\textbf{1},\textbf{2})_{\frac{1}{2}}$  & $2354$ & & \\
&&  None  & $1140$ \\
&&  $4 \times $ ($(\textbf{1},\textbf{1})_{0}$$(\textbf{1},\textbf{2})_{-\frac{1}{2}}$$(\textbf{1},\textbf{2})_{\frac{1}{2}}$),  & $430$ \\
&&  $5 \times $ ($(\textbf{1},\textbf{1})_{0}$$(\textbf{1},\textbf{2})_{\frac{1}{2}}$$(\textbf{1},\textbf{2})_{-\frac{1}{2}}$),  & $300$ \\
&&  $3 \times $ ($(\textbf{1},\textbf{1})_{0}$$(\textbf{1},\textbf{2})_{\frac{1}{2}}$$(\textbf{1},\textbf{2})_{-\frac{1}{2}}$),  & $120$ \\
&&  $1 \times $ ($(\textbf{1},\textbf{1})_{0}$$(\textbf{1},\textbf{2})_{\frac{1}{2}}$$(\textbf{1},\textbf{2})_{-\frac{1}{2}}$),  & $120$ \\
&&  $10 \times $ ($(\textbf{1},\textbf{1})_{0}$$(\textbf{1},\textbf{2})_{\frac{1}{2}}$$(\textbf{1},\textbf{2})_{-\frac{1}{2}}$),  & $94$ \\
\hline \end{tabular}}
\end{center}
\caption{A continuation of table \ref{table:One S couplings 1}.}
\label{table:One S couplings 2}
\end{table}
\begin{table}[htb]
\begin{center}
\scalebox{.65}{
\centering
\begin{tabular}{|c|c|c|c|} \hline
Exotic Set & Multiplicity & Perturbative $S$ Couplings & Multiplicity  \\ \hline
$(\textbf{1},\textbf{1})_{1}$ $(\textbf{1},\textbf{1})_{-1}$ $(\textbf{1},\textbf{2})_{-\frac{1}{2}}$ $(\textbf{1},\textbf{1})_{0}$ $(\textbf{1},\textbf{2})_{\frac{1}{2}}$  & $1496$ & & \\
&&  None  & $596$ \\
&&  $5 \times $ ($(\textbf{1},\textbf{1})_{0}$$(\textbf{1},\textbf{2})_{\frac{1}{2}}$$(\textbf{1},\textbf{2})_{-\frac{1}{2}}$),  & $180$ \\
&&  $3 \times $ ($(\textbf{1},\textbf{1})_{0}$$(\textbf{1},\textbf{1})_{1}$$(\textbf{1},\textbf{1})_{-1}$),  & $164$ \\
&&  $4 \times $ ($(\textbf{1},\textbf{1})_{0}$$(\textbf{1},\textbf{2})_{\frac{1}{2}}$$(\textbf{1},\textbf{2})_{-\frac{1}{2}}$),  & $120$ \\
&&  $3 \times $ ($(\textbf{1},\textbf{1})_{0}$$(\textbf{1},\textbf{2})_{\frac{1}{2}}$$(\textbf{1},\textbf{2})_{-\frac{1}{2}}$),  & $60$ \\
&&  $2 \times $ ($(\textbf{1},\textbf{1})_{0}$$(\textbf{1},\textbf{2})_{\frac{1}{2}}$$(\textbf{1},\textbf{2})_{-\frac{1}{2}}$),  & $60$ \\
\hline \end{tabular}}
\end{center}
\caption{A continuation of table \ref{table:One S couplings 1}.}
\label{table:One S couplings 3}
\end{table}
\begin{table}[htb]
\begin{center}
\scalebox{.65}{
\centering
\begin{tabular}{|c|c|c|c|} \hline
Exotic Set & Multiplicity & Perturbative $S$ Couplings & Multiplicity  \\ \hline
$(\textbf{1},\textbf{3})_{0}$ $(\textbf{1},\textbf{3})_{0}$ $(\textbf{1},\textbf{1})_{0}$ $(\textbf{1},\textbf{2})_{-\frac{1}{2}}$ $(\textbf{1},\textbf{2})_{\frac{1}{2}}$  & $1354$ & & \\
&&  None  & $580$ \\
&&  $5 \times $ ($(\textbf{1},\textbf{1})_{0}$$(\textbf{1},\textbf{2})_{\frac{1}{2}}$$(\textbf{1},\textbf{2})_{-\frac{1}{2}}$),  & $240$ \\
&&  $4 \times $ ($(\textbf{1},\textbf{1})_{0}$$(\textbf{1},\textbf{2})_{\frac{1}{2}}$$(\textbf{1},\textbf{2})_{-\frac{1}{2}}$),  & $230$ \\
&&  $10 \times $ ($(\textbf{1},\textbf{1})_{0}$$(\textbf{1},\textbf{2})_{\frac{1}{2}}$$(\textbf{1},\textbf{2})_{-\frac{1}{2}}$),  & $64$ \\
&&  $2 \times $ ($(\textbf{1},\textbf{1})_{0}$$(\textbf{1},\textbf{2})_{\frac{1}{2}}$$(\textbf{1},\textbf{2})_{-\frac{1}{2}}$),  & $60$ \\
&&  $3 \times $ ($(\textbf{1},\textbf{1})_{0}$$(\textbf{1},\textbf{2})_{-\frac{1}{2}}$$(\textbf{1},\textbf{2})_{\frac{1}{2}}$),  & $60$ \\
\hline \end{tabular}}
\end{center}
\caption{A continuation of table \ref{table:One S couplings 1}.}
\label{table:One S couplings 4}
\end{table}
\begin{table}[htb]
\begin{center}
\scalebox{.50}{
\centering
\begin{tabular}{|c|c|c|c|} \hline
Exotic Set & Multiplicity & Perturbative $S$ Couplings & Multiplicity  \\ \hline
$(\ov{\textbf{3}},\textbf{1})_{\frac{1}{3}}$ $(\textbf{1},\textbf{3})_{0}$ $(\textbf{1},\textbf{3})_{0}$ $(\textbf{1},\textbf{1})_{0}$ $(\textbf{3},\textbf{1})_{-\frac{1}{3}}$  & $872$ & & \\
&&  None  & $527$ \\
&&  $1 \times $ ($(\textbf{1},\textbf{1})_{0}$$(\textbf{1},\textbf{2})_{\frac{1}{2}}$$(\textbf{1},\textbf{2})_{-\frac{1}{2}}$),  & $166$ \\
&&  $2 \times $ ($(\textbf{1},\textbf{1})_{0}$$(\textbf{1},\textbf{2})_{\frac{1}{2}}$$(\textbf{1},\textbf{2})_{-\frac{1}{2}}$),  & $45$ \\
&&  $4 \times $ ($(\textbf{1},\textbf{1})_{0}$$(\textbf{1},\textbf{2})_{\frac{1}{2}}$$(\textbf{1},\textbf{2})_{-\frac{1}{2}}$),  & $22$ \\
&&  $4 \times $ ($(\textbf{1},\textbf{1})_{0}$$(\ov{\textbf{3}},\textbf{1})_{\frac{1}{3}}$$(\textbf{3},\textbf{1})_{-\frac{1}{3}}$),  & $15$ \\
&&  $3 \times $ ($(\textbf{1},\textbf{1})_{0}$$(\ov{\textbf{3}},\textbf{1})_{\frac{1}{3}}$$(\textbf{3},\textbf{1})_{-\frac{1}{3}}$),  & $15$ \\
\hline \end{tabular}}
\end{center}
\caption{A continuation of table \ref{table:One S couplings 1}.}
\label{table:One S couplings 5}
\end{table}
\begin{table}[htb]
\begin{center}
\scalebox{.65}{
\centering
\begin{tabular}{|c|c|c|c|} \hline
Exotic Set & Multiplicity & Perturbative $S$ Couplings & Multiplicity  \\ \hline
$(\textbf{1},\textbf{1})_{0}$  & $674$ & & \\
&&  None  & $480$ \\
&&  $1 \times $ ($(\textbf{1},\textbf{1})_{0}$$(\textbf{1},\textbf{2})_{-\frac{1}{2}}$$(\textbf{1},\textbf{2})_{\frac{1}{2}}$),  & $144$ \\
&&  $2 \times $ ($(\textbf{1},\textbf{1})_{0}$$(\textbf{1},\textbf{2})_{\frac{1}{2}}$$(\textbf{1},\textbf{2})_{-\frac{1}{2}}$),  & $30$ \\
&&  $4 \times $ ($(\textbf{1},\textbf{1})_{0}$$(\textbf{1},\textbf{2})_{\frac{1}{2}}$$(\textbf{1},\textbf{2})_{-\frac{1}{2}}$),  & $20$ \\
$(\textbf{1},\textbf{1})_{1}$ $(\textbf{1},\textbf{1})_{-1}$ $(\textbf{1},\textbf{1})_{0}$  & $664$ & & \\
&&  None  & $320$ \\
\hline \end{tabular}}
\end{center}
\caption{A continuation of table \ref{table:One S couplings 1}.}
\label{table:One S couplings 6}
\end{table}
\begin{table}[htb]
\begin{center}
\scalebox{.65}{
\centering
\begin{tabular}{|c|c|c|c|} \hline
Exotic Set & Multiplicity & Perturbative $S$ Couplings & Multiplicity  \\ \hline$(\textbf{1},\textbf{3})_{0}$ $(\textbf{1},\textbf{3})_{0}$ $(\textbf{1},\textbf{1})_{0}$ $(\textbf{1},\textbf{1})_{1}$ $(\textbf{1},\textbf{1})_{-1}$  & $520$ & & \\
&&  None  & $240$ \\
&&  $1 \times $ ($(\textbf{1},\textbf{1})_{0}$$(\textbf{1},\textbf{2})_{\frac{1}{2}}$$(\textbf{1},\textbf{2})_{-\frac{1}{2}}$),  & $80$ \\
&&  $2 \times $ ($(\textbf{1},\textbf{1})_{0}$$(\textbf{1},\textbf{1})_{-1}$$(\textbf{1},\textbf{1})_{1}$),  & $50$ \\
&&  $4 \times $ ($(\textbf{1},\textbf{1})_{0}$$(\textbf{1},\textbf{2})_{\frac{1}{2}}$$(\textbf{1},\textbf{2})_{-\frac{1}{2}}$),  & $30$ \\
&&  $2 \times $ ($(\textbf{1},\textbf{1})_{0}$$(\textbf{1},\textbf{2})_{\frac{1}{2}}$$(\textbf{1},\textbf{2})_{-\frac{1}{2}}$),  & $30$ \\
&&  $3 \times $ ($(\textbf{1},\textbf{1})_{0}$$(\textbf{1},\textbf{1})_{-1}$$(\textbf{1},\textbf{1})_{1}$),  $1 \times $ ($(\textbf{1},\textbf{1})_{0}$$(\textbf{1},\textbf{2})_{\frac{1}{2}}$$(\textbf{1},\textbf{2})_{-\frac{1}{2}}$),  & $20$ \\
\hline \end{tabular}}
\end{center}
\caption{A continuation of table \ref{table:One S couplings 1}.}
\label{table:One S couplings 7}
\end{table}
\begin{table}[htb]
\begin{center}
\scalebox{.65}{
\centering
\begin{tabular}{|c|c|c|c|} \hline
Exotic Set & Multiplicity & Perturbative $S$ Couplings & Multiplicity  \\ \hline
$(\ov{\textbf{3}},\textbf{1})_{-\frac{2}{3}}$ $(\textbf{1},\textbf{1})_{0}$ $(\textbf{3},\textbf{1})_{\frac{2}{3}}$  & $132$ & & \\
&&  None  & $72$ \\
&&  $1 \times $ ($(\textbf{1},\textbf{1})_{0}$$(\textbf{1},\textbf{2})_{\frac{1}{2}}$$(\textbf{1},\textbf{2})_{-\frac{1}{2}}$),  & $30$ \\
&&  $4 \times $ ($(\textbf{1},\textbf{1})_{0}$$(\textbf{3},\textbf{1})_{\frac{2}{3}}$$(\ov{\textbf{3}},\textbf{1})_{-\frac{2}{3}}$),  & $20$ \\
&&  $4 \times $ ($(\textbf{1},\textbf{1})_{0}$$(\textbf{1},\textbf{2})_{\frac{1}{2}}$$(\textbf{1},\textbf{2})_{-\frac{1}{2}}$),  & $10$ \\
$(\textbf{1},\textbf{3})_{0}$ $(\textbf{1},\textbf{3})_{0}$ $(\textbf{1},\textbf{1})_{0}$ $(\textbf{3},\textbf{1})_{\frac{2}{3}}$ $(\ov{\textbf{3}},\textbf{1})_{-\frac{2}{3}}$  & $100$ & & \\
&&  None  & $40$ \\
\hline \end{tabular}}
\end{center}
\caption{A continuation of table \ref{table:One S couplings 1}.}
\label{table:One S couplings 8}
\end{table}
\begin{table}[htb]
\begin{center}
\scalebox{.65}{
\centering
\begin{tabular}{|c|c|c|c|} \hline
Exotic Set & Multiplicity & Perturbative $S$ Couplings & Multiplicity  \\ \hline
$(\ov{\textbf{3}},\textbf{1})_{\frac{1}{3}}$ $(\textbf{1},\textbf{1})_{0}$ $(\textbf{3},\textbf{1})_{-\frac{1}{3}}$ $(\textbf{1},\textbf{2})_{-\frac{3}{2}}$ $(\textbf{1},\textbf{2})_{\frac{3}{2}}$  & $4$ & & \\
&&  $4 \times $ ($(\textbf{1},\textbf{1})_{0}$$(\textbf{1},\textbf{2})_{\frac{1}{2}}$$(\textbf{1},\textbf{2})_{-\frac{1}{2}}$),  $1 \times $ ($(\textbf{1},\textbf{1})_{0}$$(\textbf{1},\textbf{2})_{-\frac{3}{2}}$$(\textbf{1},\textbf{2})_{\frac{3}{2}}$),  & $4$ \\
$(\ov{\textbf{3}},\textbf{1})_{\frac{1}{3}}$ $(\textbf{1},\textbf{1})_{0}$ $(\textbf{3},\textbf{1})_{\frac{1}{6}}$ $(\ov{\textbf{3}},\textbf{1})_{-\frac{1}{6}}$ $(\textbf{3},\textbf{1})_{-\frac{1}{3}}$  & $4$ & & \\
&&  None  & $4$ \\
$(\textbf{1},\textbf{1})_{-\frac{1}{2}}$ $(\textbf{1},\textbf{1})_{\frac{1}{2}}$ $(\textbf{1},\textbf{2})_{\frac{1}{2}}$ $(\textbf{1},\textbf{2})_{-\frac{1}{2}}$ $(\textbf{1},\textbf{1})_{0}$  & $4$ & & \\
&&  None  & $4$ \\
$(\textbf{1},\textbf{1})_{1}$ $(\textbf{1},\textbf{1})_{-1}$ $(\textbf{1},\textbf{1})_{0}$ $(\textbf{1},\textbf{2})_{0}$ $(\textbf{1},\textbf{2})_{0}$  & $4$ & & \\
\hline \end{tabular}}
\end{center}
\caption{A continuation of table \ref{table:One S couplings 1}.}
\label{table:One S couplings 9}
\end{table}

\bibliographystyle{JHEP}
\bibliography{refs}

\providecommand{\href}[2]{#2}\begingroup\raggedright\begin{thebibliography}{100}

\bibitem{Seminar}
{\it { Talks by Jim Olsen and Marumi Kado, CERN, 15 Dec. 2015.}},  {\em LHC
  seminar \href{http://indico.cern.ch/event/442432/}{{\em ATLAS and CMS physics
  results from Run 2}}}.

\bibitem{CMS}
{\it {CMS note, Search for new physics in high mass diphoton events in
  proton-proton collisions at $13$ TeV}},  {\em CMS PAS EXO-15-004}.

\bibitem{ATLAS}
{\it {ATLAS note, \em Search for resonances decaying to photon pairs in 3.2
  fb$^{-1}$ of $pp$ collisions at $\sqrt{s}=13$ TeV with the ATLAS detector}},
  {\em ATLAS-CONF-2015-081}.

\bibitem{Harigaya:2015ezk}
K.~Harigaya and Y.~Nomura, {\it {Composite Models for the 750 GeV Diphoton
  Excess}},  \href{http://arxiv.org/abs/1512.04850}{{\tt arXiv:1512.04850}}.

\bibitem{Mambrini:2015wyu}
Y.~Mambrini, G.~Arcadi, and A.~Djouadi, {\it {The LHC diphoton resonance and
  dark matter}},  \href{http://arxiv.org/abs/1512.04913}{{\tt
  arXiv:1512.04913}}.

\bibitem{Backovic:2015fnp}
M.~Backovic, A.~Mariotti, and D.~Redigolo, {\it {Di-photon excess illuminates
  Dark Matter}},  \href{http://arxiv.org/abs/1512.04917}{{\tt
  arXiv:1512.04917}}.

\bibitem{Pilaftsis:2015ycr}
A.~Pilaftsis, {\it {Diphoton Signatures from Heavy Axion Decays at LHC}},
  \href{http://arxiv.org/abs/1512.04931}{{\tt arXiv:1512.04931}}.

\bibitem{Franceschini:2015kwy}
R.~Franceschini, G.~F. Giudice, J.~F. Kamenik, M.~McCullough, A.~Pomarol,
  R.~Rattazzi, M.~Redi, F.~Riva, A.~Strumia, and R.~Torre, {\it {What is the
  gamma gamma resonance at 750 GeV?}},
  \href{http://arxiv.org/abs/1512.04933}{{\tt arXiv:1512.04933}}.

\bibitem{Nakai:2015ptz}
Y.~Nakai, R.~Sato, and K.~Tobioka, {\it {Footprints of New Strong Dynamics via
  Anomaly}},  \href{http://arxiv.org/abs/1512.04924}{{\tt arXiv:1512.04924}}.

\bibitem{Buttazzo:2015txu}
D.~Buttazzo, A.~Greljo, and D.~Marzocca, {\it {Knocking on New Physics' door
  with a Scalar Resonance}},  \href{http://arxiv.org/abs/1512.04929}{{\tt
  arXiv:1512.04929}}.

\bibitem{DiChiara:2015vdm}
S.~Di~Chiara, L.~Marzola, and M.~Raidal, {\it {First interpretation of the 750
  GeV di-photon resonance at the LHC}},
  \href{http://arxiv.org/abs/1512.04939}{{\tt arXiv:1512.04939}}.

\bibitem{Higaki:2015jag}
T.~Higaki, K.~S. Jeong, N.~Kitajima, and F.~Takahashi, {\it {The QCD Axion from
  Aligned Axions and Diphoton Excess}},
  \href{http://arxiv.org/abs/1512.05295}{{\tt arXiv:1512.05295}}.

\bibitem{Knapen:2015dap}
S.~Knapen, T.~Melia, M.~Papucci, and K.~Zurek, {\it {Rays of light from the
  LHC}},  \href{http://arxiv.org/abs/1512.04928}{{\tt arXiv:1512.04928}}.

\bibitem{McDermott:2015sck}
S.~D. McDermott, P.~Meade, and H.~Ramani, {\it {Singlet Scalar Resonances and
  the Diphoton Excess}},  \href{http://arxiv.org/abs/1512.05326}{{\tt
  arXiv:1512.05326}}.

\bibitem{Ellis:2015oso}
J.~Ellis, S.~A.~R. Ellis, J.~Quevillon, V.~Sanz, and T.~You, {\it {On the
  Interpretation of a Possible $\sim 750$ GeV Particle Decaying into $\gamma
  \gamma$}},  \href{http://arxiv.org/abs/1512.05327}{{\tt arXiv:1512.05327}}.

\bibitem{Low:2015qep}
M.~Low, A.~Tesi, and L.-T. Wang, {\it {A pseudoscalar decaying to photon pairs
  in the early LHC run 2 data}},  \href{http://arxiv.org/abs/1512.05328}{{\tt
  arXiv:1512.05328}}.

\bibitem{Bellazzini:2015nxw}
B.~Bellazzini, R.~Franceschini, F.~Sala, and J.~Serra, {\it {Goldstones in
  Diphotons}},  \href{http://arxiv.org/abs/1512.05330}{{\tt arXiv:1512.05330}}.

\bibitem{Gupta:2015zzs}
R.~S. Gupta, S.~J{\" a}ger, Y.~Kats, G.~Perez, and E.~Stamou, {\it
  {Interpreting a 750 GeV Diphoton Resonance}},
  \href{http://arxiv.org/abs/1512.05332}{{\tt arXiv:1512.05332}}.

\bibitem{Petersson:2015mkr}
C.~Petersson and R.~Torre, {\it {The 750 GeV diphoton excess from the goldstino
  superpartner}},  \href{http://arxiv.org/abs/1512.05333}{{\tt
  arXiv:1512.05333}}.

\bibitem{Molinaro:2015cwg}
E.~Molinaro, F.~Sannino, and N.~Vignaroli, {\it {Minimal Composite Dynamics
  versus Axion Origin of the Diphoton excess}},
  \href{http://arxiv.org/abs/1512.05334}{{\tt arXiv:1512.05334}}.

\bibitem{Dutta:2015wqh}
B.~Dutta, Y.~Gao, T.~Ghosh, I.~Gogoladze, and T.~Li, {\it {Interpretation of
  the diphoton excess at CMS and ATLAS}},
  \href{http://arxiv.org/abs/1512.05439}{{\tt arXiv:1512.05439}}.

\bibitem{Cao:2015pto}
Q.-H. Cao, Y.~Liu, K.-P. Xie, B.~Yan, and D.-M. Zhang, {\it {A Boost Test of
  Anomalous Diphoton Resonance at the LHC}},
  \href{http://arxiv.org/abs/1512.05542}{{\tt arXiv:1512.05542}}.

\bibitem{Matsuzaki:2015che}
S.~Matsuzaki and K.~Yamawaki, {\it {750 GeV Diphoton Signal from One-Family
  Walking Technipion}},  \href{http://arxiv.org/abs/1512.05564}{{\tt
  arXiv:1512.05564}}.

\bibitem{Kobakhidze:2015ldh}
A.~Kobakhidze, F.~Wang, L.~Wu, J.~M. Yang, and M.~Zhang, {\it {LHC 750 GeV
  diphoton resonance explained as a heavy scalar in top-seesaw model}},
  \href{http://arxiv.org/abs/1512.05585}{{\tt arXiv:1512.05585}}.

\bibitem{Martinez:2015kmn}
R.~Martinez, F.~Ochoa, and C.~F. Sierra, {\it {Diphoton decay for a $750$ GeV
  scalar boson in an $U(1)'$ model}},
  \href{http://arxiv.org/abs/1512.05617}{{\tt arXiv:1512.05617}}.

\bibitem{Cox:2015ckc}
P.~Cox, A.~D. Medina, T.~S. Ray, and A.~Spray, {\it {Diphoton Excess at 750 GeV
  from a Radion in the Bulk-Higgs Scenario}},
  \href{http://arxiv.org/abs/1512.05618}{{\tt arXiv:1512.05618}}.

\bibitem{No:2015bsn}
J.~M. No, V.~Sanz, and J.~Setford, {\it {See-Saw Composite Higgses at the LHC:
  Linking Naturalness to the $750$ GeV Di-Photon Resonance}},
  \href{http://arxiv.org/abs/1512.05700}{{\tt arXiv:1512.05700}}.

\bibitem{Demidov:2015zqn}
S.~V. Demidov and D.~S. Gorbunov, {\it {On sgoldstino interpretation of the
  diphoton excess}},  \href{http://arxiv.org/abs/1512.05723}{{\tt
  arXiv:1512.05723}}.

\bibitem{Chao:2015ttq}
W.~Chao, R.~Huo, and J.-H. Yu, {\it {The Minimal Scalar-Stealth Top
  Interpretation of the Diphoton Excess}},
  \href{http://arxiv.org/abs/1512.05738}{{\tt arXiv:1512.05738}}.

\bibitem{Fichet:2015vvy}
S.~Fichet, G.~von Gersdorff, and C.~Royon, {\it {Scattering Light by Light at
  750 GeV at the LHC}},  \href{http://arxiv.org/abs/1512.05751}{{\tt
  arXiv:1512.05751}}.

\bibitem{Curtin:2015jcv}
D.~Curtin and C.~B. Verhaaren, {\it {Quirky Explanations for the Diphoton
  Excess}},  \href{http://arxiv.org/abs/1512.05753}{{\tt arXiv:1512.05753}}.

\bibitem{Bian:2015kjt}
L.~Bian, N.~Chen, D.~Liu, and J.~Shu, {\it {A hidden confining world on the 750
  GeV diphoton excess}},  \href{http://arxiv.org/abs/1512.05759}{{\tt
  arXiv:1512.05759}}.

\bibitem{Chakrabortty:2015hff}
J.~Chakrabortty, A.~Choudhury, P.~Ghosh, S.~Mondal, and T.~Srivastava, {\it
  {Di-photon resonance around 750 GeV: shedding light on the theory
  underneath}},  \href{http://arxiv.org/abs/1512.05767}{{\tt
  arXiv:1512.05767}}.

\bibitem{Agrawal:2015dbf}
P.~Agrawal, J.~Fan, B.~Heidenreich, M.~Reece, and M.~Strassler, {\it
  {Experimental Considerations Motivated by the Diphoton Excess at the LHC}},
  \href{http://arxiv.org/abs/1512.05775}{{\tt arXiv:1512.05775}}.

\bibitem{Csaki:2015vek}
C.~Csaki, J.~Hubisz, and J.~Terning, {\it {The Minimal Model of a Diphoton
  Resonance: Production without Gluon Couplings}},
  \href{http://arxiv.org/abs/1512.05776}{{\tt arXiv:1512.05776}}.

\bibitem{Ahmed:2015uqt}
A.~Ahmed, B.~M. Dillon, B.~Grzadkowski, J.~F. Gunion, and Y.~Jiang, {\it
  {Higgs-radion interpretation of 750 GeV di-photon excess at the LHC}},
  \href{http://arxiv.org/abs/1512.05771}{{\tt arXiv:1512.05771}}.

\bibitem{Falkowski:2015swt}
A.~Falkowski, O.~Slone, and T.~Volansky, {\it {Phenomenology of a 750 GeV
  Singlet}},  \href{http://arxiv.org/abs/1512.05777}{{\tt arXiv:1512.05777}}.

\bibitem{Bai:2015nbs}
Y.~Bai, J.~Berger, and R.~Lu, {\it {A 750 GeV Dark Pion: Cousin of a Dark
  G-parity-odd WIMP}},  \href{http://arxiv.org/abs/1512.05779}{{\tt
  arXiv:1512.05779}}.

\bibitem{Aloni:2015mxa}
D.~Aloni, K.~Blum, A.~Dery, A.~Efrati, and Y.~Nir, {\it {On a possible large
  width 750 GeV diphoton resonance at ATLAS and CMS}},
  \href{http://arxiv.org/abs/1512.05778}{{\tt arXiv:1512.05778}}.

\bibitem{Gabrielli:2015dhk}
E.~Gabrielli, K.~Kannike, B.~Mele, M.~Raidal, C.~Spethmann, and H.~Veermäe,
  {\it {A SUSY Inspired Simplified Model for the 750 GeV Diphoton Excess}},
  \href{http://arxiv.org/abs/1512.05961}{{\tt arXiv:1512.05961}}.

\bibitem{Benbrik:2015fyz}
R.~Benbrik, C.-H. Chen, and T.~Nomura, {\it {Higgs singlet as a diphoton
  resonance in a vector-like quark model}},
  \href{http://arxiv.org/abs/1512.06028}{{\tt arXiv:1512.06028}}.

\bibitem{Kim:2015ron}
J.~S. Kim, J.~Reuter, K.~Rolbiecki, and R.~R. de~Austri, {\it {A resonance
  without resonance: scrutinizing the diphoton excess at 750 GeV}},
  \href{http://arxiv.org/abs/1512.06083}{{\tt arXiv:1512.06083}}.

\bibitem{Alves:2015jgx}
A.~Alves, A.~G. Dias, and K.~Sinha, {\it {The 750 GeV $S$-cion: Where else
  should we look for it?}},  \href{http://arxiv.org/abs/1512.06091}{{\tt
  arXiv:1512.06091}}.

\bibitem{Megias:2015ory}
E.~Megias, O.~Pujolas, and M.~Quiros, {\it {On dilatons and the LHC diphoton
  excess}},  \href{http://arxiv.org/abs/1512.06106}{{\tt arXiv:1512.06106}}.

\bibitem{Carpenter:2015ucu}
L.~M. Carpenter, R.~Colburn, and J.~Goodman, {\it {Supersoft SUSY Models and
  the 750 GeV Diphoton Excess, Beyond Effective Operators}},
  \href{http://arxiv.org/abs/1512.06107}{{\tt arXiv:1512.06107}}.

\bibitem{Bernon:2015abk}
J.~Bernon and C.~Smith, {\it {Could the width of the diphoton anomaly signal a
  three-body decay?}},  \href{http://arxiv.org/abs/1512.06113}{{\tt
  arXiv:1512.06113}}.

\bibitem{Chao:2015nsm}
W.~Chao, {\it {Symmetries Behind the 750 GeV Diphoton Excess}},
  \href{http://arxiv.org/abs/1512.06297}{{\tt arXiv:1512.06297}}.

\bibitem{Arun:2015ubr}
M.~T. Arun and P.~Saha, {\it {Gravitons in multiply warped scenarios - at 750
  GeV and beyond}},  \href{http://arxiv.org/abs/1512.06335}{{\tt
  arXiv:1512.06335}}.

\bibitem{Han:2015cty}
C.~Han, H.~M. Lee, M.~Park, and V.~Sanz, {\it {The diphoton resonance as a
  gravity mediator of dark matter}},
  \href{http://arxiv.org/abs/1512.06376}{{\tt arXiv:1512.06376}}.

\bibitem{Chang:2015bzc}
S.~Chang, {\it {A Simple $U(1)$ Gauge Theory Explanation of the Diphoton
  Excess}},  \href{http://arxiv.org/abs/1512.06426}{{\tt arXiv:1512.06426}}.

\bibitem{Chakraborty:2015jvs}
I.~Chakraborty and A.~Kundu, {\it {Diphoton excess at 750 GeV: Singlet scalars
  confront naturalness}},  \href{http://arxiv.org/abs/1512.06508}{{\tt
  arXiv:1512.06508}}.

\bibitem{Ding:2015rxx}
R.~Ding, L.~Huang, T.~Li, and B.~Zhu, {\it {Interpreting $750$ GeV Diphoton
  Excess with R-parity Violation Supersymmetry}},
  \href{http://arxiv.org/abs/1512.06560}{{\tt arXiv:1512.06560}}.

\bibitem{Han:2015dlp}
H.~Han, S.~Wang, and S.~Zheng, {\it {Scalar Dark Matter Explanation of Diphoton
  Excess at LHC}},  \href{http://arxiv.org/abs/1512.06562}{{\tt
  arXiv:1512.06562}}.

\bibitem{Han:2015qqj}
X.-F. Han and L.~Wang, {\it {Implication of the 750 GeV diphoton resonance on
  two-Higgs-doublet model and its extensions with Higgs field}},
  \href{http://arxiv.org/abs/1512.06587}{{\tt arXiv:1512.06587}}.

\bibitem{Chang:2015sdy}
J.~Chang, K.~Cheung, and C.-T. Lu, {\it {Interpreting the 750 GeV Di-photon
  Resonance using photon-jets in Hidden-Valley-like models}},
  \href{http://arxiv.org/abs/1512.06671}{{\tt arXiv:1512.06671}}.

\bibitem{Bardhan:2015hcr}
D.~Bardhan, D.~Bhatia, A.~Chakraborty, U.~Maitra, S.~Raychaudhuri, and
  T.~Samui, {\it {Radion Candidate for the LHC Diphoton Resonance}},
  \href{http://arxiv.org/abs/1512.06674}{{\tt arXiv:1512.06674}}.

\bibitem{Antipin:2015kgh}
O.~Antipin, M.~Mojaza, and F.~Sannino, {\it {A natural Coleman-Weinberg theory
  explains the diphoton excess}},  \href{http://arxiv.org/abs/1512.06708}{{\tt
  arXiv:1512.06708}}.

\bibitem{Wang:2015kuj}
F.~Wang, L.~Wu, J.~M. Yang, and M.~Zhang, {\it {750 GeV Diphoton Resonance, 125
  GeV Higgs and Muon g-2 Anomaly in Deflected Anomaly Mediation SUSY Breaking
  Scenario}},  \href{http://arxiv.org/abs/1512.06715}{{\tt arXiv:1512.06715}}.

\bibitem{Cao:2015twy}
J.~Cao, C.~Han, L.~Shang, W.~Su, J.~M. Yang, and Y.~Zhang, {\it {Interpreting
  the 750 GeV diphoton excess by the singlet extension of the Manohar-Wise
  Model}},  \href{http://arxiv.org/abs/1512.06728}{{\tt arXiv:1512.06728}}.

\bibitem{Huang:2015evq}
F.~P. Huang, C.~S. Li, Z.~L. Liu, and Y.~Wang, {\it {750 GeV Diphoton Excess
  from Cascade Decay}},  \href{http://arxiv.org/abs/1512.06732}{{\tt
  arXiv:1512.06732}}.

\bibitem{Heckman:2015kqk}
J.~J. Heckman, {\it {750 GeV Diphotons from a D3-brane}},
  \href{http://arxiv.org/abs/1512.06773}{{\tt arXiv:1512.06773}}.

\bibitem{Dhuria:2015ufo}
M.~Dhuria and G.~Goswami, {\it {Perturbativity, vacuum stability and inflation
  in the light of 750 GeV diphoton excess}},
  \href{http://arxiv.org/abs/1512.06782}{{\tt arXiv:1512.06782}}.

\bibitem{Bi:2015uqd}
X.-J. Bi, Q.-F. Xiang, P.-F. Yin, and Z.-H. Yu, {\it {The 750 GeV diphoton
  excess at the LHC and dark matter constraints}},
  \href{http://arxiv.org/abs/1512.06787}{{\tt arXiv:1512.06787}}.

\bibitem{Kim:2015ksf}
J.~S. Kim, K.~Rolbiecki, and R.~R. de~Austri, {\it {Model-independent
  combination of diphoton constraints at 750 GeV}},
  \href{http://arxiv.org/abs/1512.06797}{{\tt arXiv:1512.06797}}.

\bibitem{Berthier:2015vbb}
L.~Berthier, J.~M. Cline, W.~Shepherd, and M.~Trott, {\it {Effective
  interpretations of a diphoton excess}},
  \href{http://arxiv.org/abs/1512.06799}{{\tt arXiv:1512.06799}}.

\bibitem{Cho:2015nxy}
W.~S. Cho, D.~Kim, K.~Kong, S.~H. Lim, K.~T. Matchev, J.-C. Park, and M.~Park,
  {\it {The 750 GeV Diphoton Excess May Not Imply a 750 GeV Resonance}},
  \href{http://arxiv.org/abs/1512.06824}{{\tt arXiv:1512.06824}}.

\bibitem{Cline:2015msi}
J.~M. Cline and Z.~Liu, {\it {LHC diphotons from electroweakly pair-produced
  composite pseudoscalars}},  \href{http://arxiv.org/abs/1512.06827}{{\tt
  arXiv:1512.06827}}.

\bibitem{Bauer:2015boy}
M.~Bauer and M.~Neubert, {\it {Flavor Anomalies, the Diphoton Excess and a Dark
  Matter Candidate}},  \href{http://arxiv.org/abs/1512.06828}{{\tt
  arXiv:1512.06828}}.

\bibitem{Chala:2015cev}
M.~Chala, M.~Duerr, F.~Kahlhoefer, and K.~Schmidt-Hoberg, {\it {Tricking
  Landau-Yang: How to obtain the diphoton excess from a vector resonance}},
  \href{http://arxiv.org/abs/1512.06833}{{\tt arXiv:1512.06833}}.

\bibitem{Barducci:2015gtd}
D.~Barducci, A.~Goudelis, S.~Kulkarni, and D.~Sengupta, {\it {One jet to rule
  them all: monojet constraints and invisible decays of a 750 GeV diphoton
  resonance}},  \href{http://arxiv.org/abs/1512.06842}{{\tt arXiv:1512.06842}}.

\bibitem{Boucenna:2015pav}
S.~M. Boucenna, S.~Morisi, and A.~Vicente, {\it {The LHC diphoton resonance
  from gauge symmetry}},  \href{http://arxiv.org/abs/1512.06878}{{\tt
  arXiv:1512.06878}}.

\bibitem{Murphy:2015kag}
C.~W. Murphy, {\it {Vector Leptoquarks and the 750 GeV Diphoton Resonance at
  the LHC}},  \href{http://arxiv.org/abs/1512.06976}{{\tt arXiv:1512.06976}}.

\bibitem{Feng:2015wil}
T.-F. Feng, X.-Q. Li, H.-B. Zhang, and S.-M. Zhao, {\it {The LHC 750 GeV
  diphoton excess in supersymmetry with gauged baryon and lepton numbers}},
  \href{http://arxiv.org/abs/1512.06696}{{\tt arXiv:1512.06696}}.

\bibitem{Hernandez:2015ywg}
A.~E.~C. Hernandez and I.~Nisandzic, {\it {LHC diphoton 750 GeV resonance as an
  indication of $SU(3)_c\times SU(3)_L\times U(1)_X$ gauge symmetry}},
  \href{http://arxiv.org/abs/1512.07165}{{\tt arXiv:1512.07165}}.

\bibitem{Pelaggi:2015knk}
G.~M. Pelaggi, A.~Strumia, and E.~Vigiani, {\it {Trinification can explain the
  di-photon and di-boson LHC anomalies}},
  \href{http://arxiv.org/abs/1512.07225}{{\tt arXiv:1512.07225}}.

\bibitem{Dey:2015bur}
U.~K. Dey, S.~Mohanty, and G.~Tomar, {\it {750 GeV resonance in the Dark
  Left-Right Model}},  \href{http://arxiv.org/abs/1512.07212}{{\tt
  arXiv:1512.07212}}.

\bibitem{deBlas:2015hlv}
J.~de~Blas, J.~Santiago, and R.~Vega-Morales, {\it {New vector bosons and the
  diphoton excess}},  \href{http://arxiv.org/abs/1512.07229}{{\tt
  arXiv:1512.07229}}.

\bibitem{Belyaev:2015hgo}
A.~Belyaev, G.~Cacciapaglia, H.~Cai, T.~Flacke, A.~Parolini, and H.~Serodio,
  {\it {Singlets in Composite Higgs Models in light of the LHC di-photon
  searches}},  \href{http://arxiv.org/abs/1512.07242}{{\tt arXiv:1512.07242}}.

\bibitem{Dev:2015isx}
P.~S.~B. Dev and D.~Teresi, {\it {Asymmetric Dark Matter in the Sun and the
  Diphoton Excess at the LHC}},  \href{http://arxiv.org/abs/1512.07243}{{\tt
  arXiv:1512.07243}}.

\bibitem{Jaeckel:2012yz}
J.~Jaeckel, M.~Jankowiak, and M.~Spannowsky, {\it {LHC probes the hidden
  sector}},  {\em Phys. Dark Univ.} {\bf 2} (2013) 111--117,
  [\href{http://arxiv.org/abs/1212.3620}{{\tt arXiv:1212.3620}}].

\bibitem{Cvetic:2011iq}
M.~Cveti{\v c}, J.~Halverson, and P.~Langacker, {\it {Implications of String
  Constraints for Exotic Matter and Z' s Beyond the Standard Model}},  {\em
  JHEP} {\bf 11} (2011) 058.

\bibitem{Blumenhagen:2005mu}
R.~Blumenhagen, M.~Cveti{\v c}, P.~Langacker, and G.~Shiu, {\it {Toward
  Realistic Intersecting D-Brane Models}},  {\em Ann. Rev. Nucl. Part. Sci.}
  {\bf 55} (2005) 71--139.

\bibitem{Cvetic:2016omj}
M.~Cvetic, J.~Halverson, and P.~Langacker, {\it {String Consistency, Heavy
  Exotics, and the 750 GeV Diphoton Excess at the LHC: Addendum}},
  \href{http://arxiv.org/abs/1602.06257}{{\tt arXiv:1602.06257}}.

\bibitem{Blumenhagen:2006xt}
R.~Blumenhagen, M.~Cveti{\v c}, and T.~Weigand, {\it {Spacetime instanton
  corrections in 4D string vacua: The Seesaw mechanism for D-Brane models}},
  {\em Nucl. Phys.} {\bf B771} (2007) 113--142,
  [\href{http://arxiv.org/abs/hep-th/0609191}{{\tt hep-th/0609191}}].

\bibitem{Ibanez:2006da}
L.~E. Ibanez and A.~M. Uranga, {\it {Neutrino Majorana Masses from String
  Theory Instanton Effects}},  {\em JHEP} {\bf 03} (2007) 052,
  [\href{http://arxiv.org/abs/hep-th/0609213}{{\tt hep-th/0609213}}].

\bibitem{Blumenhagen:2009qh}
R.~Blumenhagen, M.~Cveti{\v c}, S.~Kachru, and T.~Weigand, {\it {D-Brane
  Instantons in Type II Orientifolds}},  {\em Ann. Rev. Nucl. Part. Sci.} {\bf
  59} (2009) 269--296, [\href{http://arxiv.org/abs/0902.3251}{{\tt
  arXiv:0902.3251}}].

\bibitem{Cvetic:2010dz}
M.~Cveti{\v c}, J.~Halverson, and P.~Langacker, {\it {Singlet Extensions of the
  MSSM in the Quiver Landscape}},  {\em JHEP} {\bf 09} (2010) 076,
  [\href{http://arxiv.org/abs/1006.3341}{{\tt arXiv:1006.3341}}].

\bibitem{Uranga:2000xp}
A.~M. Uranga, {\it {D-Brane Probes, Rr Tadpole Cancellation and K-Theory
  Charge}},  {\em Nucl. Phys.} {\bf B598} (2001) 225--246.

\bibitem{Aldazabal:2000dg}
G.~Aldazabal, S.~Franco, L.~E. Ibanez, R.~Rabadan, and A.~M. Uranga, {\it {D =
  4 chiral string compactifications from intersecting branes}},  {\em J. Math.
  Phys.} {\bf 42} (2001) 3103--3126.

\bibitem{Ibanez:2001nd}
L.~E. Ibanez, F.~Marchesano, and R.~Rabadan, {\it {Getting just the standard
  model at intersecting branes}},  {\em JHEP} {\bf 11} (2001) 002,
  [\href{http://arxiv.org/abs/hep-th/0105155}{{\tt hep-th/0105155}}].

\bibitem{Ghilencea:2002da}
D.~M. Ghilencea, L.~E. Ibanez, N.~Irges, and F.~Quevedo, {\it {TeV scale
  Z-prime bosons from D-branes}},  {\em JHEP} {\bf 08} (2002) 016,
  [\href{http://arxiv.org/abs/hep-ph/0205083}{{\tt hep-ph/0205083}}].

\bibitem{Halverson:2013ska}
J.~Halverson, {\it {Anomaly Nucleation Constrains SU(2) Gauge Theories}},  {\em
  Phys. Rev. Lett.} {\bf 111} (2013), no.~26 261601,
  [\href{http://arxiv.org/abs/1310.1091}{{\tt arXiv:1310.1091}}].

\bibitem{Cvetic:2012kj}
M.~Cveti{\v c}, J.~Halverson, and H.~Piragua, {\it {Stringy Hidden Valleys}},
  {\em JHEP} {\bf 1302} (2013) 005.

\bibitem{Halverson:2014nwa}
J.~Halverson, N.~Orlofsky, and A.~Pierce, {\it {Vectorlike Leptons as the Tip
  of the Dark Matter Iceberg}},  {\em Phys. Rev.} {\bf D90} (2014), no.~1
  015002, [\href{http://arxiv.org/abs/1403.1592}{{\tt arXiv:1403.1592}}].

\bibitem{Cvetic:2011vz}
M.~Cveti{\v c} and J.~Halverson, {\it {TASI Lectures: Particle Physics from
  Perturbative and Non-perturbative Effects in D-braneworlds}},  in {\em
  {Proceedings, Theoretical Advanced Study Institute in Elementary Particle
  Physics (TASI 2010). String Theory and Its Applications: From meV to the
  Planck Scale: Boulder, Colorado, USA, June 1-25, 2010}}, pp.~245--292, 2011.
\newblock \href{http://arxiv.org/abs/1101.2907}{{\tt arXiv:1101.2907}}.

\bibitem{Anastasopoulos:2006cz}
P.~Anastasopoulos, M.~Bianchi, E.~Dudas, and E.~Kiritsis, {\it {Anomalies,
  anomalous U(1)'s and generalized Chern-Simons terms}},  {\em JHEP} {\bf 11}
  (2006) 057, [\href{http://arxiv.org/abs/hep-th/0605225}{{\tt
  hep-th/0605225}}].

\bibitem{Mambrini:2009ad}
Y.~Mambrini, {\it {A Clear Dark Matter gamma ray line generated by the
  Green-Schwarz mechanism}},  {\em JCAP} {\bf 0912} (2009) 005,
  [\href{http://arxiv.org/abs/0907.2918}{{\tt arXiv:0907.2918}}].

\bibitem{Cvetic:2001nr}
M.~Cveti{\v c}, G.~Shiu, and A.~M. Uranga, {\it {Chiral four-dimensional N=1
  supersymmetric type 2A orientifolds from intersecting D6 branes}},  {\em
  Nucl. Phys.} {\bf B615} (2001) 3--32,
  [\href{http://arxiv.org/abs/hep-th/0107166}{{\tt hep-th/0107166}}].

\bibitem{Anastasopoulos:2006da}
P.~Anastasopoulos, T.~Dijkstra, E.~Kiritsis, and A.~Schellekens, {\it
  {Orientifolds, hypercharge embeddings and the Standard Model}},  {\em
  Nucl.Phys.} {\bf B759} (2006) 83--146.

\bibitem{Gmeiner:2005vz}
F.~Gmeiner, R.~Blumenhagen, G.~Honecker, D.~Lust, and T.~Weigand, {\it {One in
  a billion: MSSM-like D-brane statistics}},  {\em JHEP} {\bf 01} (2006) 004,
  [\href{http://arxiv.org/abs/hep-th/0510170}{{\tt hep-th/0510170}}].

\bibitem{Djouadi:2005gi}
A.~Djouadi, {\it {The Anatomy of electro-weak symmetry breaking. I: The Higgs
  boson in the standard model}},  {\em Phys. Rept.} {\bf 457} (2008) 1--216,
  [\href{http://arxiv.org/abs/hep-ph/0503172}{{\tt hep-ph/0503172}}].

\bibitem{Huang:2015rkj}
W.-C. Huang, Y.-L.~S. Tsai, and T.-C. Yuan, {\it {Gauged Two Higgs Doublet
  Model confronts the LHC 750 GeV di-photon anomaly}},
  \href{http://arxiv.org/abs/1512.07268}{{\tt arXiv:1512.07268}}.

\bibitem{Badziak:2015zez}
M.~Badziak, {\it {Interpreting the 750 GeV diphoton excess in minimal
  extensions of Two-Higgs-Doublet models}},
  \href{http://arxiv.org/abs/1512.07497}{{\tt arXiv:1512.07497}}.

\bibitem{Cvetic:2015vit}
M.~Cveti{\v c}, J.~Halverson, and P.~Langacker, {\it {String Consistency, Heavy
  Exotics, and the $750$ GeV Diphoton Excess at the LHC}},
  \href{http://arxiv.org/abs/1512.07622}{{\tt arXiv:1512.07622}}.

\bibitem{Cheung:2015cug}
K.~Cheung, P.~Ko, J.~S. Lee, J.~Park, and P.-Y. Tseng, {\it {A Higgcision study
  on the 750 GeV Di-photon Resonance and 125 GeV SM Higgs boson with the
  Higgs-Singlet Mixing}},  \href{http://arxiv.org/abs/1512.07853}{{\tt
  arXiv:1512.07853}}.

\bibitem{Zhang:2015uuo}
J.~Zhang and S.~Zhou, {\it {Electroweak Vacuum Stability and Diphoton Excess at
  750 GeV}},  \href{http://arxiv.org/abs/1512.07889}{{\tt arXiv:1512.07889}}.

\bibitem{Hall:2015xds}
L.~J. Hall, K.~Harigaya, and Y.~Nomura, {\it {750 GeV Diphotons: Implications
  for Supersymmetric Unification}},
  \href{http://arxiv.org/abs/1512.07904}{{\tt arXiv:1512.07904}}.

\bibitem{Wang:2015omi}
F.~Wang, W.~Wang, L.~Wu, J.~M. Yang, and M.~Zhang, {\it {Interpreting 750 GeV
  Diphoton Resonance in the NMSSM with Vector-like Particles}},
  \href{http://arxiv.org/abs/1512.08434}{{\tt arXiv:1512.08434}}.

\bibitem{Salvio:2015jgu}
A.~Salvio and A.~Mazumdar, {\it {Higgs Stability and the 750 GeV Diphoton
  Excess}},  \href{http://arxiv.org/abs/1512.08184}{{\tt arXiv:1512.08184}}.

\bibitem{Son:2015vfl}
M.~Son and A.~Urbano, {\it {A new scalar resonance at 750 GeV: Towards a proof
  of concept in favor of strongly interacting theories}},
  \href{http://arxiv.org/abs/1512.08307}{{\tt arXiv:1512.08307}}.

\bibitem{Cai:2015hzc}
C.~Cai, Z.-H. Yu, and H.-H. Zhang, {\it {The 750 GeV diphoton resonance as a
  singlet scalar in an extra dimensional model}},
  \href{http://arxiv.org/abs/1512.08440}{{\tt arXiv:1512.08440}}.

\bibitem{Bizot:2015qqo}
N.~Bizot, S.~Davidson, M.~Frigerio, and J.~L. Kneur, {\it {Two Higgs doublets
  to explain the excesses $pp\rightarrow \gamma\gamma(750\ {\rm GeV})$ and $h
  \to \tau^\pm \mu^\mp$}},  \href{http://arxiv.org/abs/1512.08508}{{\tt
  arXiv:1512.08508}}.

\bibitem{Hamada:2015skp}
Y.~Hamada, T.~Noumi, S.~Sun, and G.~Shiu, {\it {An O(750) GeV Resonance and
  Inflation}},  \href{http://arxiv.org/abs/1512.08984}{{\tt arXiv:1512.08984}}.

\bibitem{Kang:2015roj}
S.~K. Kang and J.~Song, {\it {Top-phobic heavy Higgs boson as the 750 GeV
  diphoton resonance}},  \href{http://arxiv.org/abs/1512.08963}{{\tt
  arXiv:1512.08963}}.

\bibitem{Jiang:2015oms}
Y.~Jiang, Y.-Y. Li, and T.~Liu, {\it {750 GeV Resonance in the Gauged
  $U(1)'$-Extended MSSM}},  \href{http://arxiv.org/abs/1512.09127}{{\tt
  arXiv:1512.09127}}.

\bibitem{Jung:2015etr}
S.~Jung, J.~Song, and Y.~W. Yoon, {\it {How Resonance-Continuum Interference
  Changes 750 GeV Diphoton Excess: Signal Enhancement and Peak Shift}},
  \href{http://arxiv.org/abs/1601.00006}{{\tt arXiv:1601.00006}}.

\bibitem{Gu:2015lxj}
J.~Gu and Z.~Liu, {\it {Running after Diphoton}},
  \href{http://arxiv.org/abs/1512.07624}{{\tt arXiv:1512.07624}}.

\bibitem{Goertz:2015nkp}
F.~Goertz, J.~F. Kamenik, A.~Katz, and M.~Nardecchia, {\it {Indirect
  Constraints on the Scalar Di-Photon Resonance at the LHC}},
  \href{http://arxiv.org/abs/1512.08500}{{\tt arXiv:1512.08500}}.

\bibitem{Ko:2016lai}
P.~Ko, Y.~Omura, and C.~Yu, {\it {Diphoton Excess at 750 GeV in leptophobic
  U(1)$^\prime$ model inspired by $E_6$ GUT}},
  \href{http://arxiv.org/abs/1601.00586}{{\tt arXiv:1601.00586}}.

\bibitem{Palti:2016kew}
E.~Palti, {\it {Vector-Like Exotics in F-Theory and 750 GeV Diphotons}},
  \href{http://arxiv.org/abs/1601.00285}{{\tt arXiv:1601.00285}}.

\bibitem{Karozas:2016hcp}
A.~Karozas, S.~F. King, G.~K. Leontaris, and A.~K. Meadowcroft, {\it {750 GeV
  Diphoton excess from $E_6$ in F-theory GUTs}},
  \href{http://arxiv.org/abs/1601.00640}{{\tt arXiv:1601.00640}}.

\bibitem{Bhattacharya:2016lyg}
S.~Bhattacharya, S.~Patra, N.~Sahoo, and N.~Sahu, {\it {750 GeV Di-photon
  excess at CERN LHC from a dark sector assisted scalar decay}},
  \href{http://arxiv.org/abs/1601.01569}{{\tt arXiv:1601.01569}}.

\bibitem{Cao:2016udb}
J.~Cao, L.~Shang, W.~Su, Y.~Zhang, and J.~Zhu, {\it {Interpreting the 750 GeV
  diphoton excess in the Minimal Dilaton Model}},
  \href{http://arxiv.org/abs/1601.02570}{{\tt arXiv:1601.02570}}.

\bibitem{Faraggi:2016xnm}
A.~E. Faraggi and J.~Rizos, {\it {The 750 GeV diphoton LHC excess and Extra Z's
  in Heterotic-String Derived Models}},
  \href{http://arxiv.org/abs/1601.03604}{{\tt arXiv:1601.03604}}.

\bibitem{Han:2016bvl}
X.-F. Han, L.~Wang, and J.~M. Yang, {\it {An extension of two-Higgs-doublet
  model and the excesses of 750 GeV diphoton, muon g-2 and $h\to\mu\tau$}},
  \href{http://arxiv.org/abs/1601.04954}{{\tt arXiv:1601.04954}}.

\bibitem{Kawamura:2016idj}
J.~Kawamura and Y.~Omura, {\it {Diphoton excess at 750 GeV and LHC constraints
  in models with vector-like particles}},
  \href{http://arxiv.org/abs/1601.07396}{{\tt arXiv:1601.07396}}.

\bibitem{King:2016wep}
S.~F. King and R.~Nevzorov, {\it {750 GeV Diphoton Resonance from Singlets in
  an Exceptional Supersymmetric Standard Model}},
  \href{http://arxiv.org/abs/1601.07242}{{\tt arXiv:1601.07242}}.

\bibitem{Nomura:2016rjf}
T.~Nomura and H.~Okada, {\it {Generalized Zee-Babu model with 750 GeV Diphoton
  Resonance}},  \href{http://arxiv.org/abs/1601.07339}{{\tt arXiv:1601.07339}}.

\bibitem{Harigaya:2016pnu}
K.~Harigaya and Y.~Nomura, {\it {A Composite Model for the 750 GeV Diphoton
  Excess}},  \href{http://arxiv.org/abs/1602.01092}{{\tt arXiv:1602.01092}}.

\bibitem{Han:2016fli}
C.~Han, T.~T. Yanagida, and N.~Yokozaki, {\it {Implications of the 750 GeV
  Diphoton Excess in Gaugino Mediation}},
  \href{http://arxiv.org/abs/1602.04204}{{\tt arXiv:1602.04204}}.

\bibitem{Hamada:2016vwk}
Y.~Hamada, H.~Kawai, K.~Kawana, and K.~Tsumura, {\it {Models of LHC Diphoton
  Excesses Valid up to the Planck scale}},
  \href{http://arxiv.org/abs/1602.04170}{{\tt arXiv:1602.04170}}.

\bibitem{Bae:2016xni}
K.~J. Bae, M.~Endo, K.~Hamaguchi, and T.~Moroi, {\it {Diphoton Excess and
  Running Couplings}},  \href{http://arxiv.org/abs/1602.03653}{{\tt
  arXiv:1602.03653}}.

\bibitem{Martin:1997ns}
S.~P. Martin, {\it {A Supersymmetry primer}},
  \href{http://arxiv.org/abs/hep-ph/9709356}{{\tt hep-ph/9709356}}. [Adv. Ser.
  Direct. High Energy Phys.18,1(1998)].

\bibitem{Cvetic:2003ch}
M.~Cveti{\v c} and I.~Papadimitriou, {\it {Conformal field theory couplings for
  intersecting D-branes on orientifolds}},  {\em Phys. Rev.} {\bf D68} (2003)
  046001, [\href{http://arxiv.org/abs/hep-th/0303083}{{\tt hep-th/0303083}}].
  [Erratum: Phys. Rev.D70,029903(2004)].

\bibitem{Cremades:2003qj}
D.~Cremades, L.~E. Ibanez, and F.~Marchesano, {\it {Yukawa couplings in
  intersecting D-brane models}},  {\em JHEP} {\bf 07} (2003) 038,
  [\href{http://arxiv.org/abs/hep-th/0302105}{{\tt hep-th/0302105}}].

\bibitem{Lust:2004cx}
D.~Lust, P.~Mayr, R.~Richter, and S.~Stieberger, {\it {Scattering of gauge,
  matter, and moduli fields from intersecting branes}},  {\em Nucl. Phys.} {\bf
  B696} (2004) 205--250, [\href{http://arxiv.org/abs/hep-th/0404134}{{\tt
  hep-th/0404134}}].

\bibitem{Cremades:2004wa}
D.~Cremades, L.~E. Ibanez, and F.~Marchesano, {\it {Computing Yukawa couplings
  from magnetized extra dimensions}},  {\em JHEP} {\bf 05} (2004) 079,
  [\href{http://arxiv.org/abs/hep-th/0404229}{{\tt hep-th/0404229}}].

\bibitem{Cvetic:1985fp}
M.~Cveti{\v c} and C.~R. Preitschopf, {\it {Heavy Families and $N=1$
  Supergravity Within the Standard Model}},  {\em Nucl. Phys.} {\bf B272}
  (1986) 490.

\bibitem{Khachatryan:2015gza}
{\bf CMS} Collaboration, V.~Khachatryan et~al., {\it {Search for pair-produced
  vector-like B quarks in proton-proton collisions at $\sqrt{s}$ = 8 TeV}},
  \href{http://arxiv.org/abs/1507.07129}{{\tt arXiv:1507.07129}}.

\bibitem{Aad:2015kqa}
{\bf ATLAS} Collaboration, G.~Aad et~al., {\it {Search for production of
  vector-like quark pairs and of four top quarks in the lepton-plus-jets final
  state in $pp$ collisions at $\sqrt{s}=8$ TeV with the ATLAS detector}},  {\em
  JHEP} {\bf 08} (2015) 105, [\href{http://arxiv.org/abs/1505.04306}{{\tt
  arXiv:1505.04306}}].

\bibitem{Khachatryan:2015oba}
{\bf CMS} Collaboration, V.~Khachatryan et~al., {\it {Search for vector-like
  charge 2/3 T quarks in proton-proton collisions at $\sqrt(s)$ = 8 TeV}},
  {\em Phys. Rev.} {\bf D93} (2016), no.~1 012003,
  [\href{http://arxiv.org/abs/1509.04177}{{\tt arXiv:1509.04177}}].

\bibitem{Agashe:2014kda}
{\bf Particle Data Group} Collaboration, K.~A. Olive et~al., {\it {Review of
  Particle Physics (RPP)}},  {\em Chin.Phys.} {\bf C38} (2014) 090001. {\tt
  http://pdg.lbl.gov}.

\bibitem{Aad:2015dha}
{\bf ATLAS} Collaboration, G.~Aad et~al., {\it {Search for heavy lepton
  resonances decaying to a $Z$ boson and a lepton in $pp$ collisions at
  $\sqrt{s}=8$ TeV with the ATLAS detector}},  {\em JHEP} {\bf 09} (2015) 108,
  [\href{http://arxiv.org/abs/1506.01291}{{\tt arXiv:1506.01291}}].

\end{thebibliography}\endgroup

\end{document}